\documentclass[draftcls,onecolumn,11pt]{IEEEtran}

\usepackage{cite}      
\usepackage{graphicx}  
\usepackage{psfrag}    
\usepackage{subfigure} 
\usepackage{url}       
\usepackage{amssymb,amsmath,amscd,amsfonts,amstext}
\usepackage{array}
\usepackage{multirow}

\newcommand{\RR}{{\mathbb{R}}}

\newcommand{\CC}{{\mathbb{C}}}
\newcommand{\EE}{{\mathbb{E}}}
\newcommand{\T}{{\mathrm{T}}}
\newcommand{\tr}{{\mathrm{tr}}}
\newcommand{\bs}{\boldsymbol}
\newcommand{\diag}{{\mathrm{diag}}}
\newtheorem{lemma}{Lemma}
\newtheorem{corollary}{Corollary}
\newtheorem{proposition}{Proposition}
\newtheorem{theorem}{Theorem}

\newtheorem{remark}{Remark}
\def\adots{
  \mathinner{\mkern1mu\raise1pt\hbox{.}\mkern2mu\raise4pt\hbox{.}
  \mkern2mu\raise7pt\vbox{\kern7pt\hbox{.}}\mkern1mu}}
\def\build#1_#2^#3{\mathrel{
\mathop{\kern 0pt#1}\limits_{#2}^{#3}}}
\newcommand{\leftrownorm}{\left|\!\left|\!\left|}
\newcommand{\rightrownorm}{\right|\!\right|\!\right|}

\IEEEoverridecommandlockouts 

\begin{document}

\title{A Central Limit Theorem for the SINR at the LMMSE Estimator Output 
for Large Dimensional Signals} 

\author{Abla Kammoun$^{(1)}$\thanks{$(1)$ Telecom ParisTech (ENST), Paris, 
France.
\texttt{abla.kammoun@enst.fr}}, 
Malika Kharouf$^{(2)}$\thanks{$(2)$ Telecom ParisTech (ENST) and 
Casablanca University, Morocco.
\texttt{malika.kharouf@enst.fr}}, 
Walid Hachem$^{(3)}$ and Jamal Najim$^{(3)}$\thanks{$(3)$ CNRS / ENST, Paris,
France. \texttt{walid.hachem, jamal.najim@enst.fr} }
}

\markboth{A Central Limit Theorem for the SINR at the LMMSE Estimator Output}
{Kammoun \emph{et.al.}}

\maketitle

\begin{abstract}
This paper is devoted to the performance study of the 
Linear Minimum Mean Squared Error estimator for multidimensional 
signals in the large dimension regime. 
Such an estimator is frequently encountered in wireless communications
and in array processing, and the Signal to Interference and Noise Ratio 
(SINR) at its output is a popular performance index. The SINR can be modeled
as a random quadratic form which can be studied with the help of large 
random matrix theory, if one
assumes that the dimension of the received and transmitted signals
go to infinity at the same pace. This paper considers 
the asymptotic behavior of the SINR for a wide class of multidimensional
signal models that includes general multi-antenna as well as spread spectrum 
transmission models. \\ 
The expression of the deterministic approximation of the SINR in the large 
dimension regime is recalled and the SINR fluctuations around this
deterministic approximation are studied. These
fluctuations are shown to converge in distribution to the Gaussian law
in the large dimension regime, and their variance is shown to decrease as 
the inverse of the signal dimension. 
\end{abstract}

\begin{keywords}
Antenna Arrays, CDMA, Central Limit Theorem, LMMSE, Martingales, 
MC-CDMA, MIMO, Random Matrix Theory.
\end{keywords}

\section{Introduction} 
\label{sec-intro} 
Large Random Matrix Theory (LRMT) is a powerful mathematical tool 
used to study the performance of multi-user and multi-access
communication systems such as  Multiple
Input Multiple Output (MIMO) digital wireless systems, antenna arrays
for source detection and localization, spread spectrum communication
systems as Code Division Multiple Access (CDMA) and Multi-Carrier CDMA
(MC-CDMA) systems.  In most of these communication systems, the $N$
dimensional received random vector ${\bf r} \in \CC^N$ is described by
the model
\begin{equation} 
\label{eq-transmission-model} 
{\bf r} = {\bs \Sigma}{\bf s} + {\bf n}
\end{equation} 
where ${\bf s} = [s_0,s_1,\ldots,s_K]^\T$ is the unknown random vector
of transmitted symbols with size $K+1$ satisfying $\EE {\bf ss}^* =
{\bf I}_{K+1}$, the noise ${\bf n}$ is an independent Additive White
Gaussian Noise (AWGN) with covariance matrix $\EE {\bf nn}^* = \rho
{\bf I}_N$ whose variance $\rho > 0$ is known, and matrix ${\bs
  \Sigma}$ represents the known ``channel'' in the wide sense whose
structure depends on the particular system under study.  One typical
problem addressed by LRMT concerns the estimation performance by the
receiver of a given transmitted symbol, say $s_0$.

In this paper we focus on one of the most popular estimators, namely
the linear Wiener estimator, also called LMMSE for Linear Minimum Mean
Squared Error estimator: the LMMSE estimate $\hat s_0 = {\bf g}^* {\bf r}$ 
of signal $s_0$ is the one for which the $N \times 1$ vector
${\bf g}$ minimizes $\EE| \hat s_0 - s_0 |^2$. 
If we partition the channel matrix as ${\bs \Sigma} = 
\left[ {\bf y} \ {\bf Y} \right]$ where ${\bf y}$ is the first column of 
${\bs \Sigma}$ and where matrix ${\bf Y}$ has dimensions $N \times K$, then 
it is well known that vector ${\bf g}$ is given by ${\bf g} = 
\left( {\bs \Sigma} {\bs \Sigma}^* + \rho {\bf I}_N \right)^{-1} {\bf y}$. 
Usually, the performance of this estimator is evaluated
in terms of the Signal to Interference plus Noise Ratio (SINR) at its
output. Writing the received vector ${\bf r}$ as ${\bf r} = s_0 {\bf y}
+ {\bf r}_{\text{in}}$ where $s_0 {\bf y}$ is the relevant term and
${\bf r}_{\text{in}}$ represents the so-called interference plus noise
term, the SINR is given by $\beta_K = | {\bf g}^* {\bf y} |^2 /
\EE | {\bf g}^* {\bf r}_{\text{in}} |^2$. Plugging the expression of
${\bf g}$ given above into this expression, one can prove that the
SINR $\beta_K$ is given by the well-known expression:
\begin{equation} 
\label{eq-sinr-general} 
\beta_K =
{\bf y}^* \left( {\bf YY}^* + \rho {\bf I}_N \right)^{-1} {\bf y} \ .
\end{equation} 
In general, this expression does not provide a clear insight on the
impact of the channel model parameters (such as the load factor
$KN^{-1}$, the power distribution of the transmission data streams, or the
correlation structure of the channel paths in the context of
multi-antenna transmissions) on the performance of the LMMSE
estimator. 

An alternative approach, justified by the fluctuating nature of the
channel paths in the context of MIMO communications and by the
pseudo-random nature of the spreading sequences in spread spectrum
applications consists to model matrix ${\bs \Sigma}$ as a random
matrix (in this case, $\beta_K$ becomes a random SINR). The simplest
random matrix model for ${\bs \Sigma}$, corresponding to the most
canonical MIMO or CDMA transmission channels, corresponds to
independent and identically distributed (i.i.d.) entries with mean
zero and variance $N^{-1}$. In that case, LRMT shows that when
$K\to\infty$ and the load factor $KN^{-1}$ converges to a limiting
load factor $\alpha > 0$, the SINR $\beta_K$ converges almost surely
(a.s.)  to an explicit deterministic quantity
$\overline{\beta}(\alpha, \rho)$ which simply depends on the limiting
load factor $\alpha$ and on the noise variance $\rho$. As a result,
the impact of these two parameters on the LMMSE performance can be
easily evaluated \cite{tse-han-it99, ver-sha-it99}.

The LMMSE SINR large dimensional behavior for more sophisticated random 
matrix models has also been thoroughly studied 
(cf. \cite{tse-han-it99,eva-tse-it00,
  big-cai-tar-tcom00, pho-hon-tcom02, cha-hac-lou-it04,
  li-tul-ver-it04, tul-li-ver-it05, pea-col-hon-it06}) and it has been
proved that there exists a deterministic sequence $(\overline{\beta}_K)$,
generally defined as the solution of an implicit equation, such that
$\beta_K - \overline{\beta}_K \to 0$ almost surely as $K \to \infty$ and 
$\frac KN$ remains bounded away from zero and from infinity.

Beyond the convergence $\beta_K - \overline{\beta}_K \to 0$, a natural
question arises concerning the accuracy of $\overline{\beta}_K$ for
finite values of $K$. A first answer to this question consists in
evaluating the Mean Squared Error (MSE) of the SINR $\EE | \beta_K -
\overline{\beta}_K |^2$ for large $K$. A further problem is the
computation of outage probability, that is the probability for
$\beta_K-\overline{\beta}_K$ to be below a certain level. Both
problems can be addressed by establishing a Central Limit Theorem
(CLT) for $\beta_K - \overline{\beta}_K$. In this paper, we establish
such a CLT (Theorem \ref{th-clt} below) for a large class of random
matrices ${\bs \Sigma}$. We prove that there exists a sequence
$\Theta_K^2 = {\cal O}(1)$ such that $\frac{\sqrt{K}}{\Theta_K} ( \beta_K -
\overline{\beta}_K ) $ converges in distribution to the
standard normal law ${\cal N}(0,1)$ in the asymptotic regime. One can
therefore infer that the MSE asymptotically behaves like $\frac{\Theta_K^2}K$ and that the outage probability can be simply approximated by a
Gaussian tail function.

The class of random matrices ${\bs \Sigma}$ we consider in this paper
is described by the following statistical model: Assume that
\begin{equation} 
\label{eq-modele-sigma} 
{\bs \Sigma} = \big( {\Sigma}_{nk} \big)_{n=1,k=0}^{N,K} =  
\left( \frac{\sigma_{nk}}{\sqrt{K}} W_{nk} \right)_{n=1,k=0}^{N,K} 
\end{equation} 
where the complex random variables $W_{nk}$ are i.i.d. with $\EE
W_{nk} = 0$, $\EE W_{nk}^2 = 0$ and $\EE | W_{nk} |^2 = 1$ and where
$(\sigma_{nk}^2;\ 1\le n \le N;\ 0\le k \le K)$ is an array of real
numbers. Due to the fact that $\EE | {\Sigma}_{nk} |^2 = \frac{\sigma_{nk}^2}K
$, the array $(\sigma_{nk}^2)$ is referred to as a variance
profile. An important particular case is when $\sigma_{nk}^2$ is {\it separable}, that is, writes:
\begin{equation} 
\label{eq-sigma-separable} 
\sigma_{nk}^2 = d_n \tilde d_k\ ,
\end{equation}
where $(d_1, \ldots, d_N)$ and 
$(\tilde d_0, \ldots, \tilde d_{K} )$ are two vectors of real positive numbers. 

\subsection*{Applicative contexts.} 
Among the applicative contexts where the channel is described appropriately by 
model \eqref{eq-modele-sigma} or by its particular case 
\eqref{eq-sigma-separable}, let us mention: 
\begin{itemize}
\item 
Multiple antenna transmissions with $K+1$ distant sources sending their signals 
toward an array of $N$ antennas. The corresponding transmission model is 
${\bf r} = {\bs \Xi}{\bf s} + {\bf n}$ where 
${\bs \Xi} = \frac{1}{\sqrt{K}}{\bf H} {\bf P}^{1/2}$, matrix ${\bf H}$ is a $N \times (K+1)$ 
random matrix with complex Gaussian elements representing the radio channel, 
${\bf P} = \diag(p_0, \ldots, p_K)$ is the (deterministic) matrix of the powers given to the
different sources, and ${\bf n}$ is the usual AWGN satisfying 
$\EE {\bf nn}^* = \rho {\bf I}_N$. Write 
${\bf H} = \left[ {\bf h}_0 \ \cdots \ {\bf h}_K \right]$, and
assume that the columns ${\bf h}_k$ are independent, which is realistic
when the sources are distant one from another. Let ${\bf C}_k$ be the covariance
matrix ${\bf C}_k = \EE {\bf h}_k {\bf h}_k^*$ and let ${\bf C}_k = {\bf U}_k 
{\bs \Lambda}_k {\bf U}_k$ be a spectral decomposition of ${\bf C}_k$ 
where ${\bs \Lambda}_k = \diag(\lambda_{nk};\ 1 \le n \le N)$ is the
matrix of eigenvalues. Assume now that the eigenvector matrices ${\bf
  U}_0, \ldots, {\bf U}_K$ are all equal (to some matrix {\bf U},
for instance), a case considered in e.g.  \cite{kot-say-sp04} (note that
sometimes they are all identified with the Fourier $N \times N$ matrix
\cite{say-sp02}). Let ${\bs \Sigma} = {\bf U}^* {\bs \Xi}$. Then matrix ${\bs
  \Sigma}$ is described by the statistical model
\eqref{eq-modele-sigma} where the $W_{nk}$ are standard Gaussian i.i.d.,
and $\sigma^2_{nk} = \lambda_{nk} p_k$. If we partition ${\bs \Xi}$ as
${\bs \Xi} = [ {\bf x} \ {\bf X} ]$ similarly to the partition ${\bs
  \Sigma} = [ {\bf y} \ {\bf Y} ]$ above, then the SINR $\beta$ at the
output of the LMMSE estimator for the first element of vector ${\bf
  s}$ in the transmission model ${\bf r} = {\bs \Xi}{\bf s} + {\bf n}$
is
$$
\beta = 
{\bf x}^* \left( {\bf XX}^* + \rho {\bf I}_N \right)^{-1} {\bf x} = 
{\bf y}^* \left( {\bf YY}^* + \rho {\bf I}_N \right)^{-1} {\bf y} 
$$
due to the fact that ${\bf U}$ is a unitary matrix. 
Therefore, the problem of LMMSE SINR convergence 
for this MIMO model is a particular case of the general problem of convergence of the
right-hand member of \eqref{eq-sinr-general} for model \eqref{eq-modele-sigma}. 

It is also worth to say a few words about the particular case
\eqref{eq-sigma-separable} in this context.  If we assume that 
${\bs \Lambda}_0 = \cdots = {\bs \Lambda}_K$ and these matrices are
equal to ${\bs \Lambda}= \diag(\lambda_1, \ldots, \lambda_N)$, then the 
model for ${\bf H}$ is the well-known Kronecker model with correlations at 
reception \cite{shi-fos-gan-kah-tcom00}.
In this case, 
\begin{equation}
\label{eq-mimo-separable}
{\bs \Sigma} = {\bf U}^* {\bs \Xi} = 
\frac{1}{\sqrt{K}} {\bf U}^* {\bf H} {\bf P}^{1/2} = 
\frac{1}{\sqrt{K}} {\bs \Lambda}^{1/2} {\bf W} {\bf P}^{1/2} 
\end{equation} 
where ${\bf W}$ is a random matrix with iid standard Gaussian elements.
This model coincides with the separable variance profile model 
\eqref{eq-sigma-separable} with $d_n = \lambda_{n}$ and $\tilde d_k = p_k$.

\item CDMA transmissions on flat fading channels. 
Here $N$ is the spreading factor, $K+1$ is the number of users, and  
\begin{equation}
\label{eq-cdma-flat} 
{\bs \Sigma} = {\bf V} {\bf P}^{1/2} 
\end{equation} 
where ${\bf V}$ is the $N \times (K+1)$ signature matrix assumed here to have
random i.i.d. elements with mean zero and variance $N^{-1}$, and 
where ${\bf P} = \diag(p_0, \ldots, p_K)$ is the users powers matrix. 
In this case, the variance profile is separable with $d_n = 1$ and 
$\tilde d_k = \frac KN p_k$. Note that elements of ${\bf V}$ are not Gaussian 
in general. 

\item 
Cellular MC-CDMA transmissions on frequency selective channels. In the uplink 
direction, the matrix ${\bs \Sigma}$ is written as:
\begin{equation}
\label{eq-mccdma-uplink} 
{\bs \Sigma} = \left[ {\bf H}_0 {\bf v}_0 \ \cdots \ 
{\bf H}_{K+1} {\bf v}_{K+1} \right] \ ,
\end{equation} 
where ${\bf H}_k = \diag(h_k(\exp(2 \imath \pi (n-1) / N);\ 1\le n \le N)$ 
is the radio channel matrix of user $k$ ($\imath= \sqrt{-1}$) in the
discrete Fourier domain (here $N$ is the number of frequency bins)
and ${\bf V} = [ {\bf v}_0, \cdots, {\bf v}_K
]$ is the $N \times (K+1)$ signature matrix with i.i.d. elements 
as in the CDMA case above. Modeling this time the channel
transfer functions as deterministic functions, we have $\sigma^2_{nk}
= \frac KN | h_k(\exp(2 \imath \pi (n-1) / N)) |^2$. \\
In the downlink direction, we have
\begin{equation}
\label{eq-mccdma-downlink} 
{\bs \Sigma} = {\bf H} {\bf V} {\bf P}^{1/2} 
\end{equation}
where ${\bf H} = \diag(h(\exp(2 \imath \pi (n-1) / N);\ 1\le n \le N)$ is the
radio channel matrix in the discrete Fourier domain, the $N \times (K+1)$ 
signature matrix ${\bf V}$ is as above, and 
${\bf P} = \diag(p_0, \ldots, p_K)$ is the matrix of the powers given to the
different users. Model \eqref{eq-mccdma-downlink} coincides with the separable
variance profile model \eqref{eq-sigma-separable} with 
$d_n = \frac KN | h(\exp(2 \imath \pi (n-1) / N)) |^2$ and $d_k = p_k$. 
\end{itemize} 

\subsection*{About the literature.} 
The asymptotic approximation $\overline\beta_K$ (first order result)
is connected with the asymptotic eigenvalue distribution of Gram matrices
${\bf Y} {\bf Y}^*$ where elements of ${\bf Y}$ are
described by the model \eqref{eq-modele-sigma}, and can be found in
the mathematical LRMT literature in the work of Girko \cite{gir-90}
(see also \cite{sil-bai-jma95} and \cite{shl-96}).  Applications in
the field of wireless communications can be found in e.g.
\cite{cha-hac-lou-it04} in the separable case and in
\cite{tul-li-ver-it05} in the general variance profile case. 

Concerning the CLT for $\beta_K - \overline\beta_K$ (second order
result), only some particular cases of the general model
\eqref{eq-modele-sigma} have been considered in the literature among which 
the i.i.d. case ($\sigma^2_{nk} = 1$) is studied in \cite{tse-zei-it00} 
(and based on
a result of \cite{sil-ap90} pertaining to the asymptotic
behavior of the eigenvectors of ${\bf Y} {\bf Y}^*$).  The more
general CDMA model \eqref{eq-cdma-flat} has been considered in
\cite{pan-guo-zhou-aap07}, using a result of \cite{got-tik-02}. The
model used in this paper includes the models of \cite{tse-zei-it00}
and \cite{pan-guo-zhou-aap07} as particular cases. 

Fluctuations of other performance indexes such as Shannon's mutual
information 
$\mathbb{E} \log \det \left( \frac{\Sigma \Sigma^*}{\rho} + {\bf I}_N\right)$ 
have also been studied at length. Let us cite \cite{HKLNP06pre} where the CLT 
is established in the separable case and \cite{hac-lou-naj-(clt)-(sub)aap07} 
for a CLT in the general variance profile case. Similar results concerning the 
mutual information are found in \cite{mou-sim-sen-it03} and in 
\cite{mou-sim-it07}. 

\subsection*{Limiting expressions vs $K$-dependent expressions.} 

As one may check in Theorems \ref{th-approx-beta} and 
\ref{th-clt} below, we deliberately chose to provide deterministic expressions 
$\overline\beta_K$ and $\Theta_K^2$ which remain bounded but do not 
necessarily converge as $K\to\infty$.  
For instance, Theorem \ref{th-approx-beta} only states that 
$\beta_K - \overline\beta_K \to 0$ almost surely. No conditions which would
guarantee the convergence of $\beta_K$ are added. 
This approach has two advantages: $1)$ such expressions for 
$\overline\beta_K$ and $\Theta_K^2$ exist for very general variance profiles 
$(\sigma_{nk}^2)$ while limiting expressions may not, and $2)$ they provide a natural discretization which can 
easily be implemented. 

The statements about these deterministic approximations are valid 
within the following asymptotic regime:
\begin{equation}\label{asympto}
K \to \infty,\quad \liminf \frac KN > 0\quad \textrm{and}\quad 
\limsup \frac KN < \infty\ .  
\end{equation}
Note that $\frac KN$ is not required to converge. In the remainder of the paper, the
notation ``$K\to\infty$'' will refer to \eqref{asympto}.

We note that in the particular case where $\frac K N \to \alpha > 0$ and the 
variance profile is obtained by a regular sampling of a continuous 
function $f$ i.e. 
$\sigma_{nk}^2 = f\left(\frac nN, \frac k{K+1}\right)$, it is possible to prove that 
$\overline\beta_K$ and $\Theta_K^2$ converge towards limits that can be 
characterized by integral equations. 

\subsection*{Principle of the approach.} 
The approach used here is simple and powerful. It is based on the 
approximation of $\beta_K$ by the sum of a martingale difference sequence and 
on the use of the CLT for martingales \cite{bil-PM-livre95}. We note that 
apart from the LRMT context, such a technique has been used recently in 
\cite{bha-gir-kok-spl07} to establish a CLT on general quadratic forms of the 
type ${\bf z}^* {\bf A} {\bf z}$ where ${\bf A}$ is a deterministic matrix 
and ${\bf z}$ is a random vector with i.i.d. elements. 

\subsection*{Paper organization.} 

In Section \ref{sec-first-order}, first-order results, whose
presentation and understanding is compulsory to state the CLT, are
recalled. The CLT, which is the main contribution of this paper, is
provided in Section \ref{sec-clt}. In Section \ref{sec-simus}, 
simulations and numerical illustrations are provided.
The proof of the main theorem (Theorem \ref{th-clt}) in given in 
Section \ref{sec-proof} while the Appendix gathers proofs of 
intermediate results.


\subsection*{Notations.} 
Given a complex $N \times N$ matrix ${\bf X} = [ x_{ij} ]_{i,j=1}^N$, 
denote by $\| {\bf X} \|$ its spectral norm, and by 
$\leftrownorm {\bf X} \rightrownorm_\infty$ its maximum row sum norm,
i.e., $\leftrownorm {\bf X} \rightrownorm_\infty = 
\max_{1\le i \le N} \sum_{j=1}^N | x_{ij} |$.  
Denote by $\| \cdot \|$ the Euclidean norm of a vector and by 
$\| \cdot \|_\infty$ its $\max$ (or $\ell_\infty$) norm.  

\section{First Order Results: The SINR Deterministic Approximation} 
\label{sec-first-order} 

In the sequel, we shall often show explicitly the dependence on $K$ in the
notations. Consider the quadratic form \eqref{eq-sinr-general}: 
$$\beta_K =
{\bf y}^* \left( {\bf YY}^* + \rho {\bf I}_N \right)^{-1} {\bf y}\ ,
$$  
where the sequence of matrices ${\bs \Sigma}(K) = [ {\bf y}(K) \ {\bf Y}(K) ]$ 
is given by 
$$ 
{\bs \Sigma}(K)\ =\ \left( {\Sigma}_{nk}(K) \right)_{n=1,k=0}^{N,K}\ =\  
 \left( \frac{\sigma_{nk}(K)}{\sqrt{K}} W_{nk} \right)_{n=1,k=0}^{N,K} \ .
$$
Let us state the main assumptions:
\paragraph*{{\bf A1}} The complex random variables 
$( W_{nk};\ n \ge 1,\ k \ge 0 )$ are i.i.d. with 
$\EE W_{10} = 0$, $\EE W_{10}^2 = 0$, $\EE | W_{10} |^2 = 1$ and 
$\EE | W_{10} |^8 < \infty$.

\paragraph*{{\bf A2}} There exists a real number $\sigma_{\max} < \infty$ 
such that
$$ 
\sup_{K\ge 1}\
\max_{\genfrac{}{}{0pt}{}{1\le n\le N}{0\le k\le K} }
|\sigma_{nk}(K)| \leq \sigma_{\max} \ . 
$$

Let $(a_m; 1\le m\le M)$ be complex numbers, then $\diag(a_m; 1\le
m\le M)$ refers to the $M\times M$ diagonal matrix whose diagonal
elements are the $a_m$'s. If ${\bf A}=(a_{ij})$ is a square matrix, then
$\diag({\bf A})$ refers to the matrix $\diag(a_{ii})$. Consider the
following diagonal matrices based on the variance profile along the
columns and the rows of ${\bs \Sigma}$:
\begin{equation} 
\begin{array}{lcl} 
{\bf D}_{k}(K) &=& \diag(\sigma^2_{1k}(K), \cdots, \sigma^2_{Nk}(K)),  \quad
0\le k\le K \\
\widetilde{\bf D}_{n}(K) &=& \diag(\sigma^2_{n1}(K), \cdots, 
\sigma^2_{nK}(K)),  \quad 1\le n\le  N . 
\end{array} 
\label{eq-def-D} 
\end{equation} 
\paragraph*{{\bf A3}} 
The variance profile satisfies 
$$
\liminf_{K \geq 1} \min_{0\le k\le K} \frac 1K \tr {\bf D}_k(K) > 0 \ . 
$$

Since $\EE | W_{10} |^2 = 1$, one has $\EE | W_{10} |^4 \ge 1$. The following is needed:
\paragraph*{{\bf A4}} 
At least one of the following conditions is satisfied: 
$$
\EE| W_{10} |^4 > 1 \quad \text{or} \quad 
\liminf_K 
\frac{1}{K^2} \tr\left( {\bf D}_0(K) \sum_{k=1}^K {\bf D}_k(K) \right) 
> 0 \ . 
$$

\begin{remark}
If needed, one can attenuate the assumption on the eighth moment in 
{\bf A1}. For instance, one can adapt without difficulty the proofs in this 
paper to the case where $\EE | W_{10} |^{4+\epsilon} < \infty$ for 
$\varepsilon > 0$. We assumed $\EE | W_{10} |^8 < \infty$ because at 
some places we rely on results of \cite{hac-lou-naj-(clt)-(sub)aap07} which
are stated with the assumption on the eighth moment. \\ 
Assumption {\bf A3} is technical. It has already 
appeared in \cite{hac-lou-naj-aap07}. \\
Assumption {\bf A4} is necessary to get a non-vanishing variance $\Theta_K^2$ 
in Theorem \ref{th-clt}. 
\end{remark}
\vspace*{0.05\columnwidth} 
The following definitions will be of help in the sequel. A complex
function $t(z)$ belongs to class ${\cal S}$ if $t(z)$ is analytical in
the upper half plane $\CC_+ = \{ z \in \CC \ ; \ \text{im}(z) > 0\}$,
if $t(z) \in \CC_+$ for all $z \in \CC_+$ and if $\text{im}(z) | t(z) |$ is 
bounded over the upper half plane $\CC_+$. 

Denote by ${\bf Q}_K(z)$ and $\widetilde{\bf Q}_K(z)$ the resolvents of 
${\bf Y}(K){\bf Y}(K)^*$ and ${\bf Y}(K)^* {\bf Y}(K)$ respectively, that is the 
$N \times N$ and $K \times K$ matrices defined by:
$$
{\bf Q}_K(z) = ( {\bf Y}(K) {\bf Y}(K)^* - z {\bf I}_N )^{-1}\qquad \textrm{and}\qquad 
\widetilde{\bf Q}_K(z) = 
( {\bf Y}(K)^* {\bf Y}(K) - z {\bf I}_K )^{-1}\ .
$$

\subsection{The SINR Deterministic approximation}

It is known \cite{gir-90, hac-lou-naj-aap07} that there exists a
deterministic diagonal $N \times N$ matrix function ${\bf T}(z)$ that
approximates the resolvent ${\bf Q}(z)$ in the following sense: Given a test
matrix ${\bf S}$ with bounded spectral norm, the quantity $\frac 1K
\tr\ {\bf S} ({\bf Q}(z) - {\bf T}(z))$ converges a.s.~to zero as
$K\to\infty$. It is also known that the approximation
$\overline\beta_K$ of the SINR $\beta_K$ is simply related to ${\bf
  T}(z)$ (cf. Theorem \ref{th-approx-beta}). As we shall see, matrix ${\bf T}(z)$
also plays a fundamental role in the second order result (Theorem \ref{th-clt}).

In the following theorem, we recall the definition and some of the
main properties of ${\bf T}(z)$.

\begin{theorem} The following hold true: 
\label{th-T-Q} 
\begin{enumerate} 
\item 
\label{th-1st-order-system} 
\cite[Theorem 2.4]{hac-lou-naj-aap07} 
Let $(\sigma_{nk}^2(K);\ 1\le n \le N;\ 1\le k \le K)$ be a sequence
of arrays of real numbers and consider the matrices ${\bf D}_{k}(K)$ and 
$\widetilde{\bf D}_{n}(K)$ defined in \eqref{eq-def-D}. 
The system of $N+K$ functional equations
\begin{equation}
\label{eq-systeme-T} 
\left\{\begin{array}{lcl}
t_{n,K}(z) &=& \displaystyle{\frac{-1}
{z\left( 1 + \frac 1K \tr (\widetilde{\bf D}_{n}(K)  
\widetilde{\bf T}_K(z))\right)},
\quad 1\leq n \leq N} \\
\tilde{t}_{k,K}(z) &=& \displaystyle{\frac{-1}
{z\left( 1 + \frac 1K \tr ({\bf D}_{k}(K) {\bf T}_K(z))\right)},
\quad 1\leq k \leq K}
\end{array} \right.
\end{equation} 
where 
$$
{\bf T}_K(z) = \diag(t_{1,K}(z), \ldots, t_{N,K}(z)), \quad 
\widetilde{\bf T}_K(z) = \diag(\tilde t_{1,K}(z), \ldots, 
\tilde t_{K,K}(z))  
$$
admits a unique solution $({\bf T}, \widetilde{\bf T})$ among the diagonal 
matrices for which the $t_{n,K}$'s and the $\tilde t_{k,K}$'s belong to class
${\cal S}$. Moreover, 
functions $t_{n,K}(z)$ and $\tilde{t}_{k,K}(z)$ admit an analytical
continuation over $\CC - \RR_+$ which is real and positive for 
$z \in (-\infty, 0)$. 
\item 
\label{th-1st-order-convergence} 
\cite[Theorem 2.5]{hac-lou-naj-aap07} 
Assume that Assumptions {\bf A1} and {\bf A2} hold true.
Consider the sequence of random matrices ${\bf Y}(K){\bf Y}(K)^*$ where ${\bf Y}$ has dimensions $N\times K$ and 
whose entries are given by
${\bf Y}_{nk} = \frac{\sigma_{nk}}{\sqrt{K}}  W_{nk}$.
For every sequence ${\bf S}_K$ of $N \times N$ diagonal matrices
and every sequence $\widetilde{\bf S}_K$ of $K \times K$ diagonal matrices
with 
$$
\sup_K \max\left( \| {\bf S}_K \|,  
\| \widetilde{\bf S}_K \| \right) < \infty \, ,
$$
the following limits hold true almost surely:
\begin{eqnarray*}
\lim_{K\to\infty} \frac 1K \tr\ {\bf S}_K \left( {\bf Q}_K(z) - {\bf T}_K(z) 
\right)  
&=& 0, \qquad \forall z \in \CC - \RR_+, \\
\lim_{K\to\infty}
\frac 1K \tr\ \widetilde{\bf S}_K 
\left( \widetilde{\bf Q}_K(z) - \widetilde{\bf T}_K(z) \right) &=& 0, \qquad \forall z \in \CC - \RR_+ \ .
\end{eqnarray*}

\end{enumerate} 
\end{theorem} 
\vspace*{0.05\columnwidth} 

The following lemma which reproduces \cite[Lemma 2.7]{bai-sil-anpro98} will 
be used throughout the paper.
It characterizes the asymptotic behavior of an important class of quadratic 
forms:
\begin{lemma} 
\label{lm-jack} 
Let ${\bf x} = [X_1, \ldots, X_N]^\T$ be a $N\times 1$ vector 
where the $X_n$ are centered i.i.d. complex random variables with unit variance. 
Let ${\bf A}$ be a deterministic $N \times N$ complex matrix. 
Then, for any $p \geq 2$, there exists a constant $C_p$ depending on $p$
only such that  
\begin{equation}
\label{eq-jack-longue} 
\EE\left| \frac 1N {\bf x}^* {\bf A} {\bf x} 
- \frac 1N \tr({\bf A}) \right|^p 
\leq 
\frac{C_p}{N^{p}} 
\left( \left( \EE|X_1|^4 \tr({\bf AA}^*) \right)^{p/2} + 
\EE|X_1|^{2p} \tr\left( ({\bf AA}^*)^{p/2} \right) \right) \ . 
\end{equation} 
\end{lemma} 
\vspace*{0.05\columnwidth} 
Noticing that $\tr({\bf AA}^*) \le N \| {\bf A} \|^2$ and that 
$\tr\left( ({\bf AA}^*)^{p/2} \right) \le N \| {\bf A} \|^p$, we obtain the
simpler inequality 
\begin{equation}
\label{eq-lm-jack-simple} 
\EE\left| \frac 1N {\bf x}^* {\bf A} {\bf x} 
- \frac 1N \tr({\bf A}) \right|^p 
\leq 
\frac{C_p}{N^{p/2}} \| {\bf A} \|^p 
\left( \left( \EE|X_1|^4 \right)^{p/2} + \EE|X_1|^{2p} \right) 
\end{equation} 
which is useful in case one has bounds on $\| {\bf A} \|$. \\ 
Using Theorem \ref{th-T-Q} and Lemma \ref{lm-jack}, we are in position to
characterize the asymptotic behavior of the quadratic form $\beta_K$
given by \eqref{eq-sinr-general}. We begin by rewriting $\beta_K$ as 
\begin{equation}
\label{eq-fq}
\beta_K = \frac 1K
{\bf w}_{0}^* {\bf D}_{0}^{1/2}
\left( {\bf YY}^* + \rho {\bf I}_N \right)^{-1} 
{\bf D}_{0}^{1/2} {\bf w}_{0}
= 
\frac 1K
{\bf w}_{0}^* {\bf D}_{0}^{1/2}
{\bf Q}(-\rho) 
{\bf D}_{0}^{1/2} {\bf w}_{0}
\end{equation}
where the $N\times 1$ vector ${\bf w}_{0}$ is given by
${\bf w}_{0} = [ W_{10}, \ldots, W_{N0} ]^\T$
and the diagonal matrix ${\bf D}_{0}$ is given by \eqref{eq-def-D}.  
Recall that ${\bf w}_{0}$ and ${\bf Q}$ are independent and that
$\| {\bf D}_0 \| \le \sigma^2_{\max}$ by {\bf A2}.  
Furthermore, one can easily notice that 
$\| {\bf Q}(-\rho) \| = \| ({\bf YY}^* + \rho {\bf I})^{-1} \| 
\le 1/\rho$. 

Denote by $\EE_{\bf Q}$ the conditional expectation with respect to ${\bf Q}$, i.e. 
$\EE_{\bf Q}= \EE (\ \cdot \ \| {\bf Q})$. From Inequality \eqref{eq-lm-jack-simple}, there exists a constant $C > 0$
for which 
\begin{eqnarray*}
\EE \EE_{\bf Q} \left| \beta_K - \frac 1K \tr {\bf D}_{0} {\bf Q}(-\rho) 
\right|^4
&\leq& 
\frac{C}{K^2} 
\left( \frac N K \right)^2
\EE \| {\bf D}_{0} {\bf Q} \|^4
\left( (\EE|W_{10}|^4)^2 + 
\EE|W_{10}|^8 \right) \\
&\leq& 
\frac{C}{K^2}
\left( \frac N K \right)^2
\left( \frac{\sigma^2_{\max}}{\rho} \right)^4
\left( (\EE|W_{10}|^4)^{2} + 
\EE|W_{10}|^{8} \right) \\
&=& {\mathcal O}\left( \frac 1 {K^2}\right)\ .
\end{eqnarray*} 
By the Borel-Cantelli Lemma, we therefore have 
$$
\beta_K - \frac 1K \tr ({\bf D}_{0} {\bf Q}(-\rho))  
\xrightarrow[K\to\infty]{} 0 \quad \text{a.s.} 
$$
Using this result, simply apply Theorem 
\ref{th-T-Q}--(\ref{th-1st-order-convergence}) with
${\bf S} = {\bf D}_{0}$ (recall that $\| {\bf D}_{0} \| 
\leq \sigma_{\max}^2$) to obtain:
\begin{theorem} 
\label{th-approx-beta}
Let $\overline\beta_K = \displaystyle{
\frac1K \tr ( {\bf D}_{0}(K) {\bf T}_K(-\rho))} $ where ${\bf T}_K$ is
given by Theorem \ref{th-T-Q}--(\ref{th-1st-order-system}). Assume
{\bf A1} and {\bf A2}. Then 
$$
\beta_K - \overline\beta_K
\xrightarrow[K\to\infty]{} 0 \quad \text{a.s.}
$$
\end{theorem}
\vspace*{0.05\columnwidth} 
\subsection{The deterministic approximation in the separable case} 
\label{subsec-1er-ordre-separable}  
In the separable case $\sigma_{nk}(K) = d_n(K) \tilde{d}_k(K)$, matrices 
${\bf D}_{k}(K)$ and $\widetilde{\bf D}_{n}(K)$ are written as  
${\bf D}_{k}(K) = \tilde{d}_k(K) {\bf D}(K)$
and $\widetilde{\bf D}_{n}(K) = d_n(K) \widetilde{\bf D}(K)$ where 
${\bf D}(K)$ and $\widetilde{\bf D}(K)$ are the diagonal matrices 
\begin{equation} 
\label{eq-d-dtilde} 
{\bf D}(K) = \diag(d_1(K), \ldots, d_N(K)), \quad 
\widetilde{\bf D}(K) = \diag(\tilde d_1(K), \ldots, \tilde d_K(K)) \ .
\end{equation} 
and one can check that the system of $N+K$ equations leading to 
${\bf T}_K$ and $\widetilde{\bf T}_K$ simplifies into a system of two 
equations, and Theorem \ref{th-T-Q} takes the following form:
\begin{proposition}\cite[Sec.~3.2]{hac-lou-naj-aap07}
\label{prop-first-order-separable}
\begin{enumerate} 
\item 
\label{prop-delta} 
Assume $\sigma_{nk}^2(K) = d_n(K) \tilde d_k(K)$.  
Given $\rho > 0$, the system of two equations
\begin{equation}
\label{eq-first-order-separable}
\left\{\begin{array}{lcl}
\delta_K(\rho) &=&
\frac{1}{K} \tr \left(
{\bf D} \left( \rho({\bf I}_N + \tilde\delta_K(\rho) 
{\bf D}) \right)^{-1} \right)  \\
\tilde\delta_K(\rho) &=&
\frac{1}{K} \tr \left(
\widetilde{\bf D} \left( \rho({\bf I}_K + \delta_K(\rho) \widetilde{\bf D}) 
\right)^{-1} \right)
\end{array}\right.
\end{equation}
where ${\bf D}$ and $\widetilde{\bf D}$ are given by \eqref{eq-d-dtilde} 
admits a unique solution $(\delta_K(\rho), \tilde{\delta}_K(\rho))$. 
Moreover, in this case matrices ${\bf T}(-\rho)$ and 
$\widetilde{\bf T}(-\rho)$ provided by Theorem 
\ref{th-T-Q}--(\ref{th-1st-order-system}) coincide with 
\begin{equation}
\label{eq-T-Ttilde-separable} 
{\bf T}(-\rho) = \frac 1\rho ({\bf I} + \tilde\delta(\rho) {\bf D} )^{-1}
\quad \text{and} \quad 
\widetilde{\bf T}(-\rho) = \frac 1\rho 
({\bf I} + \delta(\rho) \widetilde{\bf D} )^{-1} \ . 
\end{equation} 
\item 
Assume that {\bf A1} and {\bf A2} hold true. Let matrices ${\bf S}_K$
and $\widetilde{\bf S}_K$ be as in Theorem 
\ref{th-T-Q}--(\ref{th-1st-order-convergence}). Then, almost surely 
$\frac 1K \tr\left( {\bf S}_K \left( {\bf Q}_K(-\rho) - {\bf T}_K(-\rho)
\right) \right) \to 0$ and 
$\frac 1K \tr\left( \widetilde{\bf S}_K
\left( \widetilde{\bf Q}_K(-\rho) - \widetilde{\bf T}_K(-\rho) \right) 
\right) \to 0$ as $K\to\infty$. 
\end{enumerate} 
\end{proposition}
\vspace*{0.05\columnwidth} 

With these equations we can adapt the result of Theorem \ref{th-approx-beta} 
to the separable case. Notice that ${\bf D}_0 = \tilde d_0 {\bf D}$ and that 
$\delta(\rho)$ given by the system \eqref{eq-first-order-separable} coincides 
with $\frac 1K \tr( {\bf D} {\bf T})$, hence
\begin{proposition}
\label{prop-beta-searable} 
Assume that $\sigma_{nk}^2(K) = d_n(K) \tilde d_k(K)$, and that {\bf
  A1} and {\bf A2} hold true. Then
$$
\frac{\beta_K}{\tilde d_0} - \delta_K(\rho) 
\xrightarrow[K\to\infty]{} 0 \quad \text{a.s.}
$$
where $\delta_K(\rho)$ is given by Proposition 
\ref{prop-first-order-separable}--(\ref{prop-delta}).  
\end{proposition} 
\vspace*{0.05\columnwidth} 

Let us provide a more explicit expression of $\delta_K$ which will be
used in Section \ref{sec-simus} to illustrate the SINR behavior for the MIMO
Model \eqref{eq-mimo-separable} and for MC-CDMA downlink Model
\eqref{eq-mccdma-downlink}. By combining the two equations in System 
\eqref{eq-first-order-separable}, it turns out that $\delta = \delta_K(\rho)$ 
is the unique solution of the implicit equation
\begin{equation}
\label{eq-sinr-mccdma} 
\delta = 
\frac 1K \sum_{n=0}^{N-1} 
\frac{ d_n }
{ \rho + \frac 1K d_n \sum_{k=1}^K \frac{p_k}{1 + p_k \delta} } \ . 
\end{equation} 
Recall that in the case of the MIMO model \eqref{eq-mimo-separable}, 
$d_n = \lambda_n$ and $\tilde d_k = p_k$, while in the case of the MC-CDMA
downlink model \eqref{eq-mccdma-downlink}, 
$d_n = \frac KN | h(\exp(2 \imath \pi (n-1) / N) |^2$ 
and $\tilde d_k = p_k$ again. Here $\tilde d_0 = p_0$ is the power of the user 
of interest (user $0$), and therefore $\beta_K / \tilde d_0$ is the normalized 
SINR of this user. 
Notice that $\delta_K(\rho)$ is almost the same for all users, hence 
the normalized SINRs for all users are close to each other for large $K$. 
Their common deterministic approximation is given by \eqref{eq-sinr-mccdma} 
which is the discrete analogue of the integral equation (16) in 
\cite{cha-hac-lou-it04}. \\ 
This example will be continued in Section \ref{sec-clt}.

\section{Second order results: The Central Limit Theorem}
\label{sec-clt}

The following theorem is the main result of this paper. Its proof is postponed to Section \ref{sec-proof}. 
\begin{theorem} 
\label{th-clt}
\begin{enumerate}
\item 
\label{th-clt-bornes} 
Assume that {\bf A2}, {\bf A3} and {\bf A4} hold true. 
Let ${\bf A}_K$ and ${\bs \Delta}_K$ be the $K\times K$ matrices
\begin{eqnarray}
{\bf A}_K &=&
\left[
\frac 1K \frac{\frac{1}{K} \tr {\bf D}_{\ell} {\bf D}_{m} {\bf T}(-\rho)^2}
{\left(1 + \frac{1}{K} \tr {\bf D}_{\ell} {\bf T}(-\rho) \right)^2}
\right]_{\ell,m=1}^K \quad \text{and} 
\label{eq-def-A} \\ 
{\bs \Delta}_K &=&
\diag\left( 
\left( 1 + \frac 1K \tr {\bf D}_{\ell} {\bf T}(-\rho) 
\right)^2; 1 \le \ell \le K \right)\ ,
\nonumber 
\end{eqnarray}
where ${\bf T}$ is defined in Theorem 
\ref{th-T-Q}--(\ref{th-1st-order-system}). Let ${\bf g}_K$ be the $K \times 1$ 
vector
$$
{\bf g}_K = \left[ \frac 1K \tr {\bf D}_0 {\bf D}_1 {\bf T}(-\rho)^2, \cdots,
\frac 1K \tr {\bf D}_0 {\bf D}_K {\bf T}(-\rho)^2 \right]^\T\ .
$$
Then the sequence of real numbers
\begin{equation} 
\label{eq-theta}
\Theta_K^2 =
\frac{1}{K} {\bf g}^\T ( {\bf I}_K - {\bf A})^{-1} {\bs \Delta}^{-1} {\bf g}
+ 
( \EE| W_{10} |^4 - 1 )
\frac 1K \tr {\bf D}_0^2 {\bf T}(-\rho)^2 
\end{equation} 
is well defined and furthermore
$$
0 < \liminf_K \Theta_K^2 \leq \limsup_K \Theta_K^2 < \infty \ . 
$$
\item 
\label{th-clt-cvg} 
Assume in addition {\bf A1}. Then the sequence $\beta_K = {\bf y}^*
({\bf YY}^* + \rho {\bf I})^{-1} {\bf y}$ satisfies
$$
\frac{\sqrt{K}}{\Theta_K} \left( \beta_K - \overline\beta_K\right)
\xrightarrow[K\to\infty]{} {\cal N}(0,1)
$$
in distribution where $\overline\beta_K=\frac1K \tr \, {\bf D}_{0} {\bf T}_K$ is defined in the statement of 
Theorem \ref{th-approx-beta}. 
\end{enumerate} 
\end{theorem} 

\begin{remark} {\em (Comparison with other performance indexes)}
\label{rem-decroissance} 
It is interesting to compare the ``Mean Squared Error'' (MSE) related to the 
SINR $\beta_K$: $MSE(\beta_K)= \EE( \beta_K - \overline{\beta}_K )^2$, 
with the MSE related to Shannon's mutual
information per transmit dimension $I = \frac 1 K \log\det( \rho {\bs
  \Sigma}{\bs \Sigma}^* + {\bf I})$ (studied in
\cite{mou-sim-sen-it03, hac-lou-naj-(clt)-(sub)aap07} for instance):
$$
MSE( \beta_K) \quad  \propto\quad  {\mathcal O}\left( \frac 1K \right)\qquad \textrm{while}\qquad 
MSE(I) \quad \propto \quad {\mathcal O}\left( \frac 1{K^2} \right)\ .
$$
\end{remark} 
\begin{remark} {\em (On the achievability of the minimum of the variance)}
\label{rem-psk}
Recall that the variance writes
$$
\Theta_K^2 = \frac{1}{K} {\bf g}^\T ( {\bf I}_K - {\bf A})^{-1} {\bs
  \Delta}^{-1} {\bf g} + ( \EE| W_{10} |^4 - 1 ) \frac 1K \tr\, {\bf
  D}_0^2 {\bf T}^2\ .
$$
As $\EE|W_{10}|^2 = 1$, one clearly has $\EE|W_{10}|^4 - 1 \geq 0$
with equality if and only if $|W_{10}| = 1$ with probability one.
Moreover, we shall prove in the sequel (Section
\ref{sec-proof-th-clt-bornes}) that $\liminf_K \frac 1K {\bf D}_0(K)
{\bf T}_K^2 > 0$. Therefore $( \EE| W_{10} |^4 - 1 ) \frac 1K \tr\,
{\bf D}_0^2 {\bf T}^2$ is nonnegative, and is zero if and only if
$|W_{10}| = 1$ with probability one. As a consequence, $\Theta_K^2$ is
minimum with respect to the distribution of the $W_{nk}$ if and only
if these random variables have their values on the unit circle.  In
the context of CDMA and MC-CDMA, this is the case when the signature
matrix elements are elements of a PSK constellation. In multi-antenna
systems, the $W_{nk}$'s are frequently considered as Gaussian which
induces a penalty on the SINR asymptotic MSE with respect to the unit
norm case.
\end{remark} 
\vspace*{0.05\columnwidth} 
In the separable case, $\Theta_K^2 = \tilde d_0^2 \Omega_K^2$ where
$\Omega_K^2$ is given by the following corollary.  
\begin{corollary}
\label{cor-clt-separable}
Assume that {\bf A2} is satisfied and 
that $\sigma^2_{nk} = d_n \tilde d_k$. Assume moreover that  
\begin{equation}\label{condition-sep}
\min\left( \liminf_K \frac 1K \tr({\bf D}(K)), 
\liminf_K \frac 1K \tr(\widetilde{\bf D}(K)) \right) > 0
\end{equation}
where ${\bf D}$ and $\widetilde{\bf D}$ are given by \eqref{eq-d-dtilde}. 
Let $\gamma = \frac 1K \tr {\bf D}^2 {\bf T}^2$ and $\tilde\gamma = \frac 1K
\tr \widetilde{\bf D}^2 \widetilde{\bf T}^2$.
Then the sequence 
\begin{equation} 
\label{eq-omega-separable} 
\Omega_K^2 =
\gamma \left(
\frac{\rho^2 \gamma \tilde\gamma}{ 1 - \rho^2 \gamma \tilde\gamma}
+ \left( \EE|W_{10} |^4 - 1 \right) \right)
\end{equation} 
satisfies $0 < \liminf_K \Omega_K^2 \leq \limsup_K \Omega_K^2 < \infty$. 
If, in addition, {\bf A1} holds true, then: 
$$
\frac{\sqrt{K}}{\Omega_K} \left(\frac{\beta_K}{\tilde d_0} -
  \delta_K\right) \xrightarrow[K\to\infty]{} {\cal N}(0,1)
$$
in distribution.
\end{corollary}

\begin{remark}
Condition \eqref{condition-sep} is the counterpart of Assumption {\bf
  A3} in the case of a separable variance profile and suffices to
establish $0 < \liminf_K (1-\rho^2\gamma \tilde \gamma) \leq \limsup_K
(1-\rho^2\gamma \tilde \gamma) < 1$ (see for instance
\cite{HKLNP06pre}), hence the fact that $0 < \liminf_K \Omega_K^2 \leq
\limsup_K \Omega_K^2 < \infty$. The remainder of the proof of
Corollary \ref{cor-clt-separable} is postponed to Appendix
\ref{anx-proof-cor-clt-separable}.
\end{remark}
\begin{remark}
As a direct application of Corollary \ref{cor-clt-separable} (to be used
in Section \ref{sec-simus} below), let us provide the expressions of 
$\gamma$ and $\tilde\gamma$ for the MIMO Model \eqref{eq-mimo-separable} or
MC-CDMA downlink Model \eqref{eq-mccdma-downlink}. 
From \eqref{eq-d-dtilde}--\eqref{eq-T-Ttilde-separable}, we get
\begin{eqnarray}
\gamma &=& 
\frac{1}{K} \sum_{n=0}^{N-1} 
\left( \frac{d_n}
{ \rho + \rho d_n \tilde\delta } \right)^2 
= 
\frac{1}{K} \sum_{n=0}^{N-1} 
\left( \frac{d_n}
{\rho + \frac 1K  d_n \sum_{k=1}^K \frac{p_k}{1 + p_k \delta}}\right)^2  
\nonumber  \\
\tilde\gamma &=& 
\frac 1K 
\sum_{k=1}^K 
\left( \frac{p_k}{\rho + \rho p_k \delta } \right)^2  
\nonumber 
\end{eqnarray}
where we recall that $d_n = \lambda_n$ for Model \eqref{eq-mimo-separable}, 
$d_n = \frac KN | h(\exp(2 \imath \pi (n-1) / N) |^2$
for Model \eqref{eq-mccdma-downlink}, and $\delta$ is the solution of 
\eqref{eq-sinr-mccdma}.  
\end{remark}


\section{Simulations}
\label{sec-simus} 

\subsection{The general (non necessarily separable) case} 

In this section, the accuracy of the Gaussian approximation is verified by 
simulation. In order to validate the results of Theorems \ref{th-approx-beta} 
and \ref{th-clt} for practical values of $K$, we consider the example of a
MC-CDMA transmission in the uplink direction. 
We recall that $K$ is the number of interfering users in this context.  
In the simulation, the discrete time channel impulse response of user $k$ is
represented by the vector with $L=5$ coefficients ${\bf g}_k = [ g_{k,0},
\ldots, g_{k,L-1} ]^\T$. In the simulations, these vectors are generated
pseudo-randomly according to the complex multivariate Gaussian law 
${\cal CN}(0, 1/L {\bf I}_L)$. Setting the number of frequency bins to $N$, 
the channel matrix ${\bf H}_k$ for user $k$ in the frequency domain
(see Eq.~\eqref{eq-mccdma-uplink}) is ${\bf H}_k = \diag( h_k(\exp(2 \imath
\pi (n-1) / N); 1 \le n \le N)$ where 
$h_k(z) = \frac{\sqrt{P_k}}{\| {\bf g}_k \|} \sum_{l=0}^{L-1} g_{k,l} z^{-l}$,
the norm $\| {\bf g}_k \|$ is the Euclidean norm of ${\bf g}_k$ 
and $P_k$ is the power received from user $k$. 
Concerning the distribution of the user powers $P_k$, we assume that these
are arranged into five power classes with powers $P, 2P, 4P, 8P$ and $16P$ 
with relative frequencies given by Table \ref{tab-powers}. 
\begin{table}[htbp]
\label{tab-powers} 
\caption{Power classes and relative frequencies}
\begin{center}
\begin{tabular}{|c|c|c|c|c|c|}
\hline
Class & 1 &2 & 3 & 4 & 5 \\
\hline
Power & $P$ & $2P$ & $4P$ & $8P$ & $16P$ \\
\hline
Relative frequency & 
${1} / {8}$ & ${1} / {4}$ & ${1} / {4}$ & ${1} / {8}$ & ${1} / {4}$\\
\hline
\end{tabular}
\end{center}
\end{table}
The user of interest (User $0$) is assumed to belong to Class $1$.
Finally, we assume that the number $K$ of interfering users is set to 
$K = N / 2$. \\ 

\begin{figure}
 \begin{center}
      \includegraphics[scale=0.6]{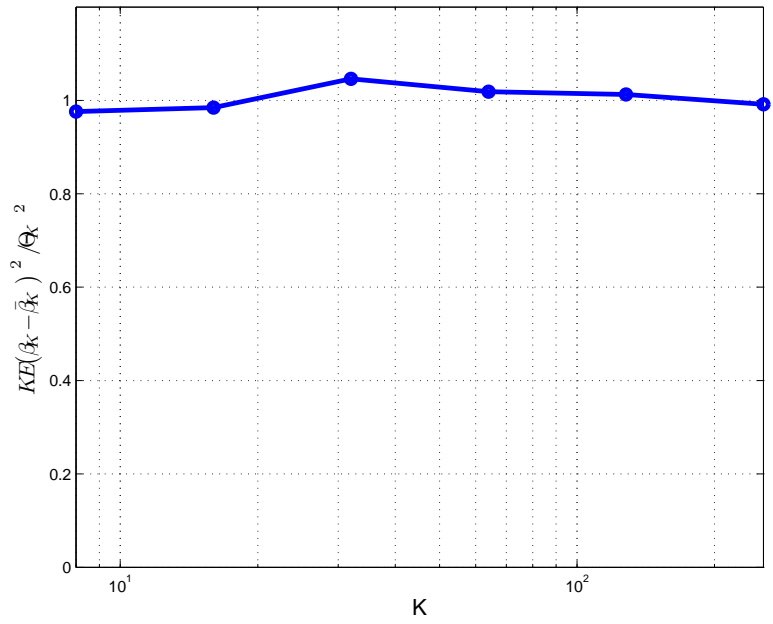}
   \end{center}
   \caption{SINR normalized MSE vs $K$}
   \label{fig:variance-vs-K}
\end{figure}

\begin{figure}
 \begin{center}
      \includegraphics[scale=0.6]{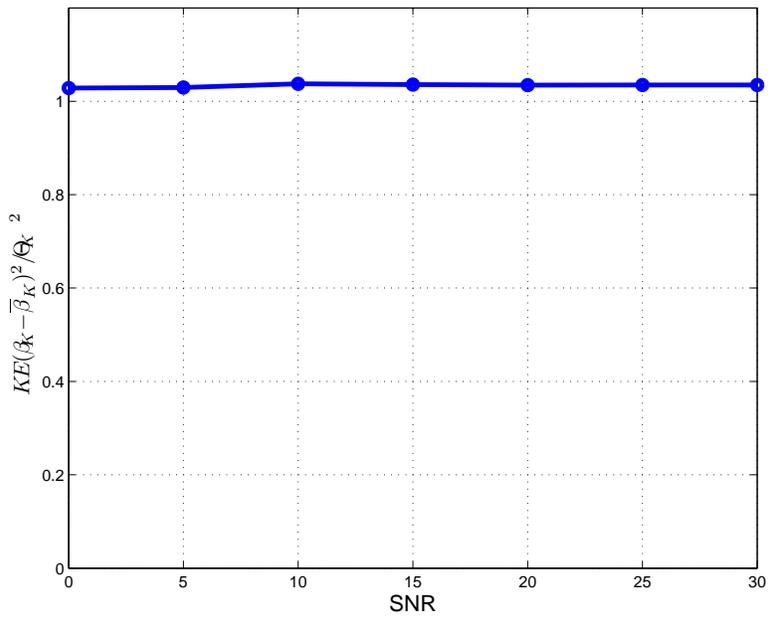}
   \end{center}
   \caption{SINR normalized MSE vs SNR}
   \label{fig:variance-vs-snr}
\end{figure}

\begin{figure}[htbp]
   \begin{center}
      \includegraphics[scale=0.6]{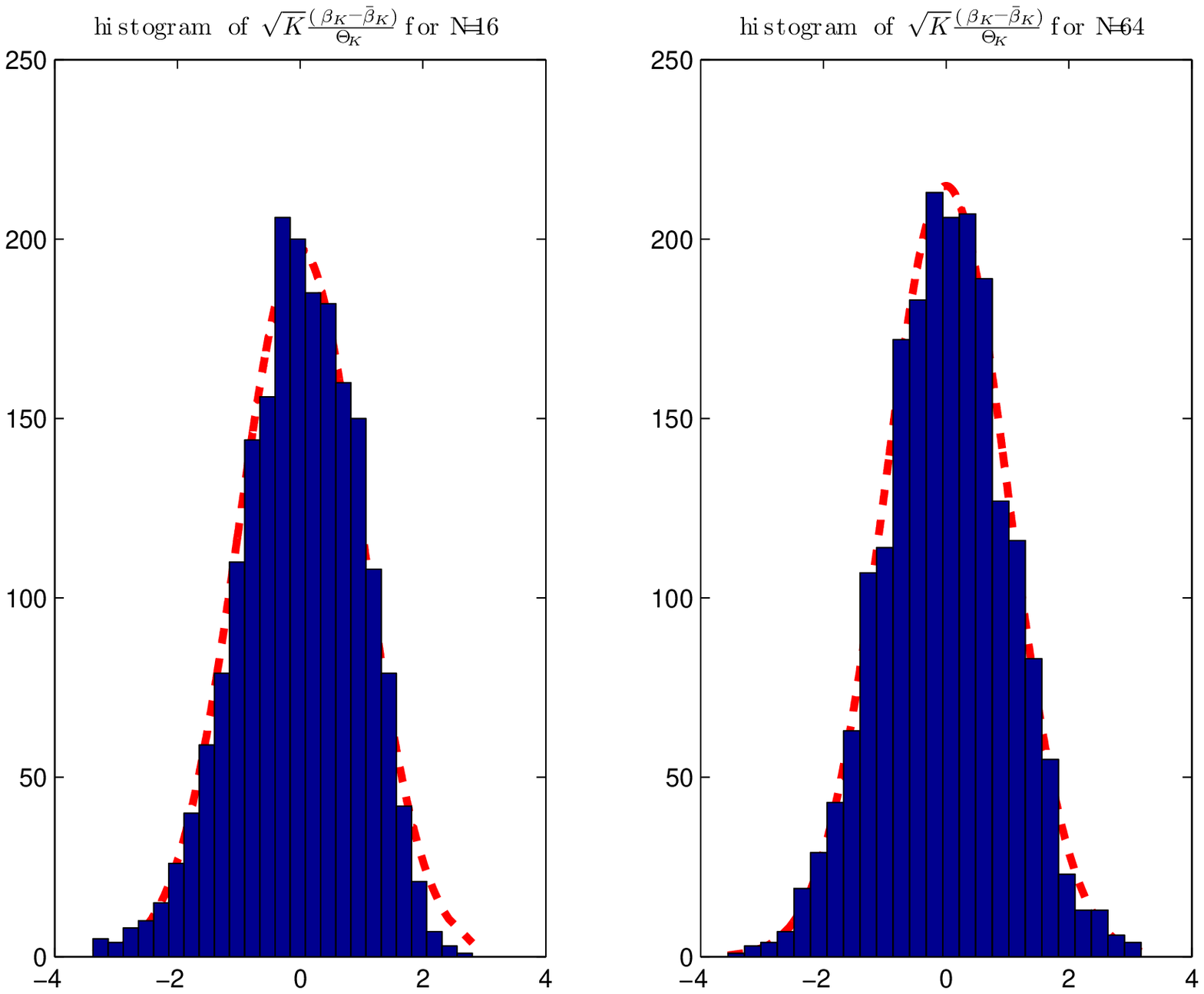}
   \end{center}
   \caption{Histogram of $\sqrt{K}(\beta_K-\bar{\beta}_K)$ 
   for $N=16$ and $N=64$.}
   \label{fig:histogram}
\end{figure}

\begin{figure}[htbp]
   \begin{center}
      \includegraphics[scale=0.6]{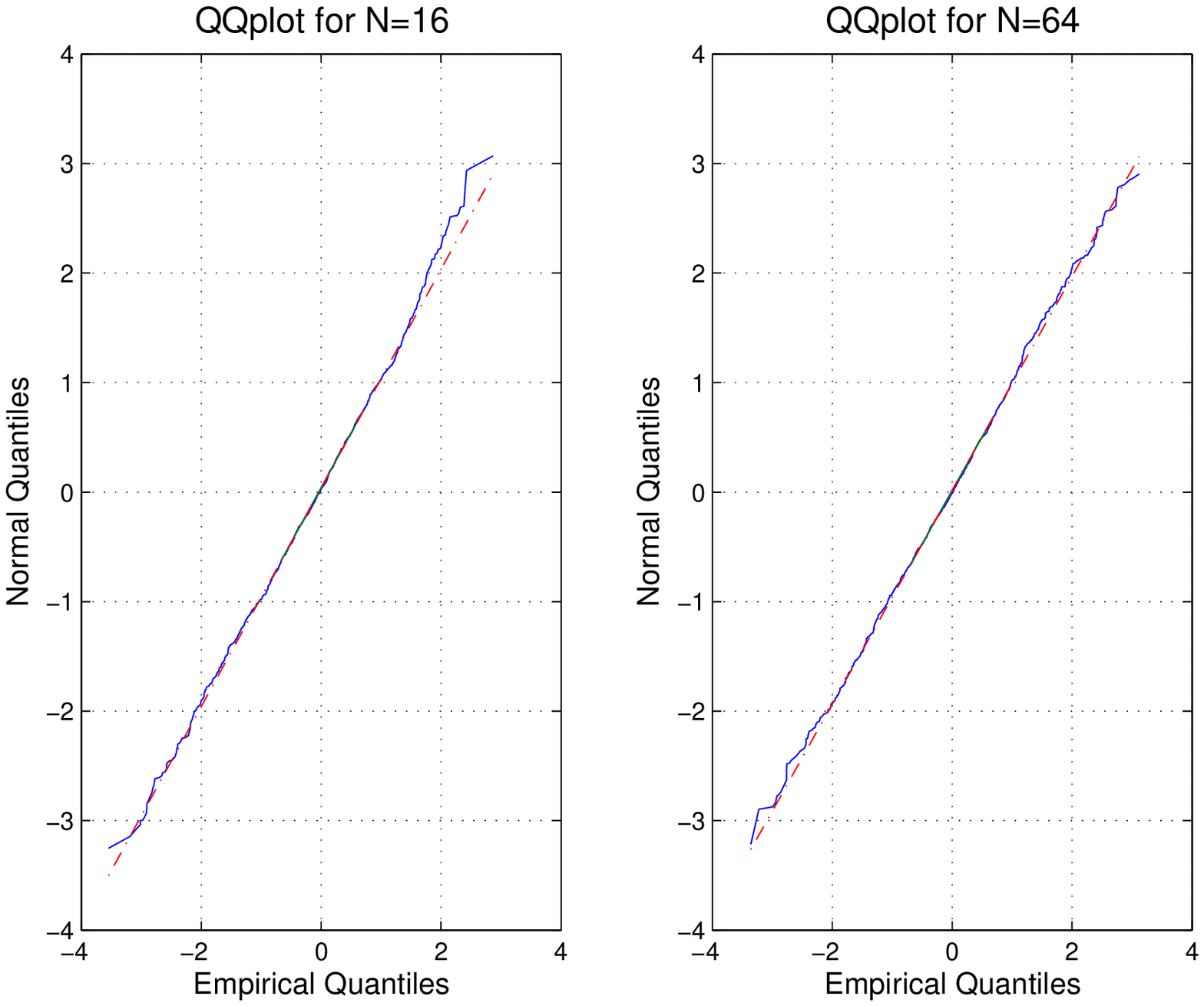}
   \end{center}
   \caption{Q-Q plot for $\sqrt{K}(\beta_K-\bar{\beta}_K)$,  
   $N=16$ and $N=64$; dash doted line is the 45 degree line.}
   \label{fig-qqtest}
\end{figure}

In Figure \ref{fig:variance-vs-K}, the Signal over Noise Ratio (SNR) 
$P / \rho$ for the user of interest is fixed to $10$ dB. 
The evolution of $K \EE (\beta_K-\bar{\beta}_K)^2 /\Theta_K^2$ for this user
(where $\EE (\beta_K-\bar{\beta}_K)^2$ is measured numerically) is shown with
respect to $K$. We note that this quantity is close to one for values of
$K$ as small as $K=8$. \\ 
In Figure \ref{fig:variance-vs-snr}, $K$ is set to $K=64$, and the 
SINR normalized MSE $K \EE (\beta_K-\bar{\beta}_K)^2 / \Theta_K^2$ is plotted 
with respect to the input SNR $P / \rho$. This figure also confirms the fact
that the MSE asymptotic approximation is highly accurate. \\
Figure \ref{fig:histogram} shows the histogram of 
$\sqrt{K}(\beta_K-\bar{\beta}_K)/\Theta_K$ for $N=16$ and $N=64$. 
This figure gives an idea of the similarity between the distribution 
of $\sqrt{K}(\beta_K-\bar{\beta}_K)/\Theta_K$ and ${\cal N}(0,1)$. \\ 
More precisely, Figure \ref{fig-qqtest} quantifies this similarity through
a Quantile-Quantile plot.  
\subsection{The separable case} 
\begin{figure*}[h]
\begin{center}
\begin{tabular}{cc}
\includegraphics[width=0.4\textwidth]{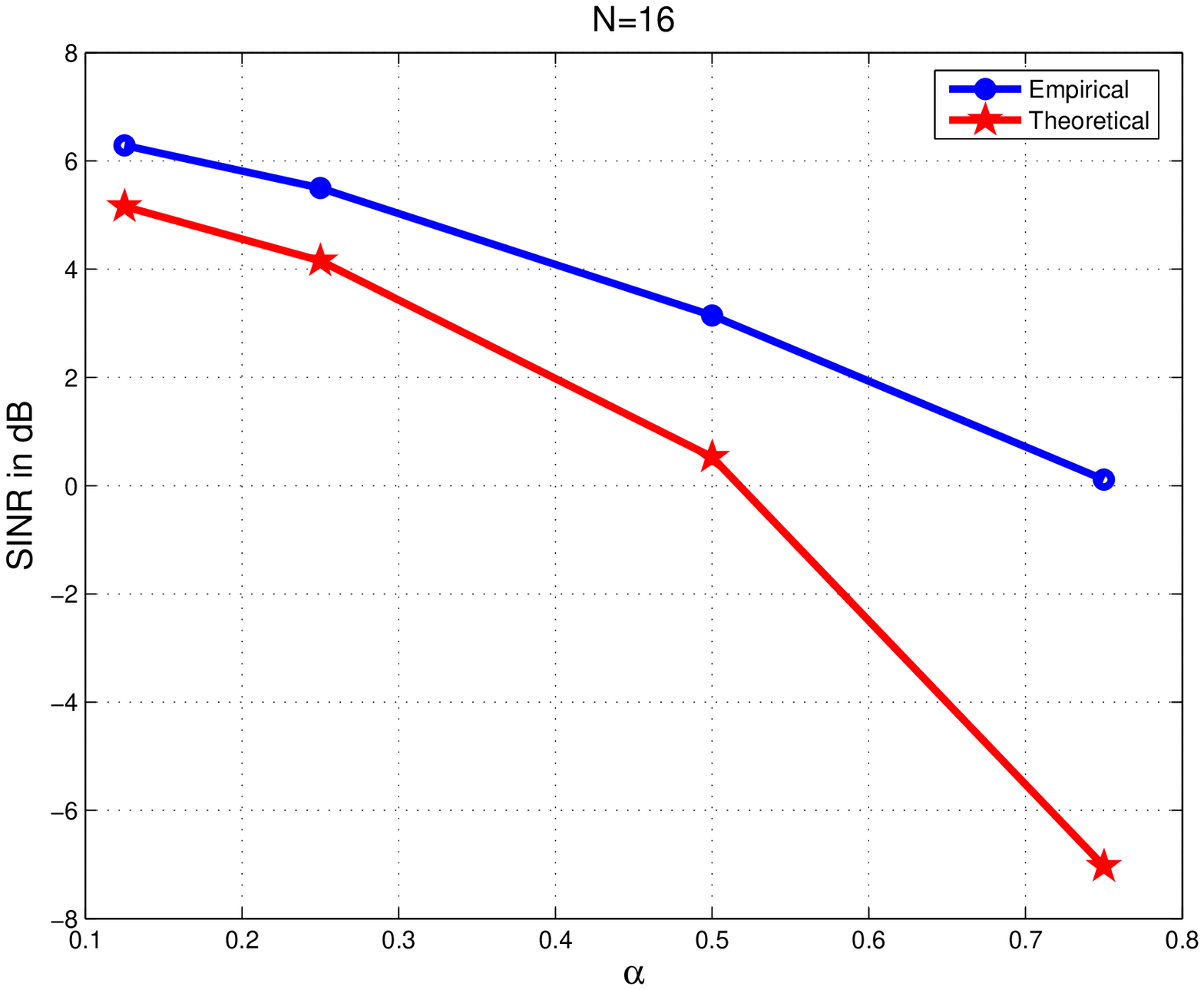}
&\includegraphics[width=0.4\textwidth]{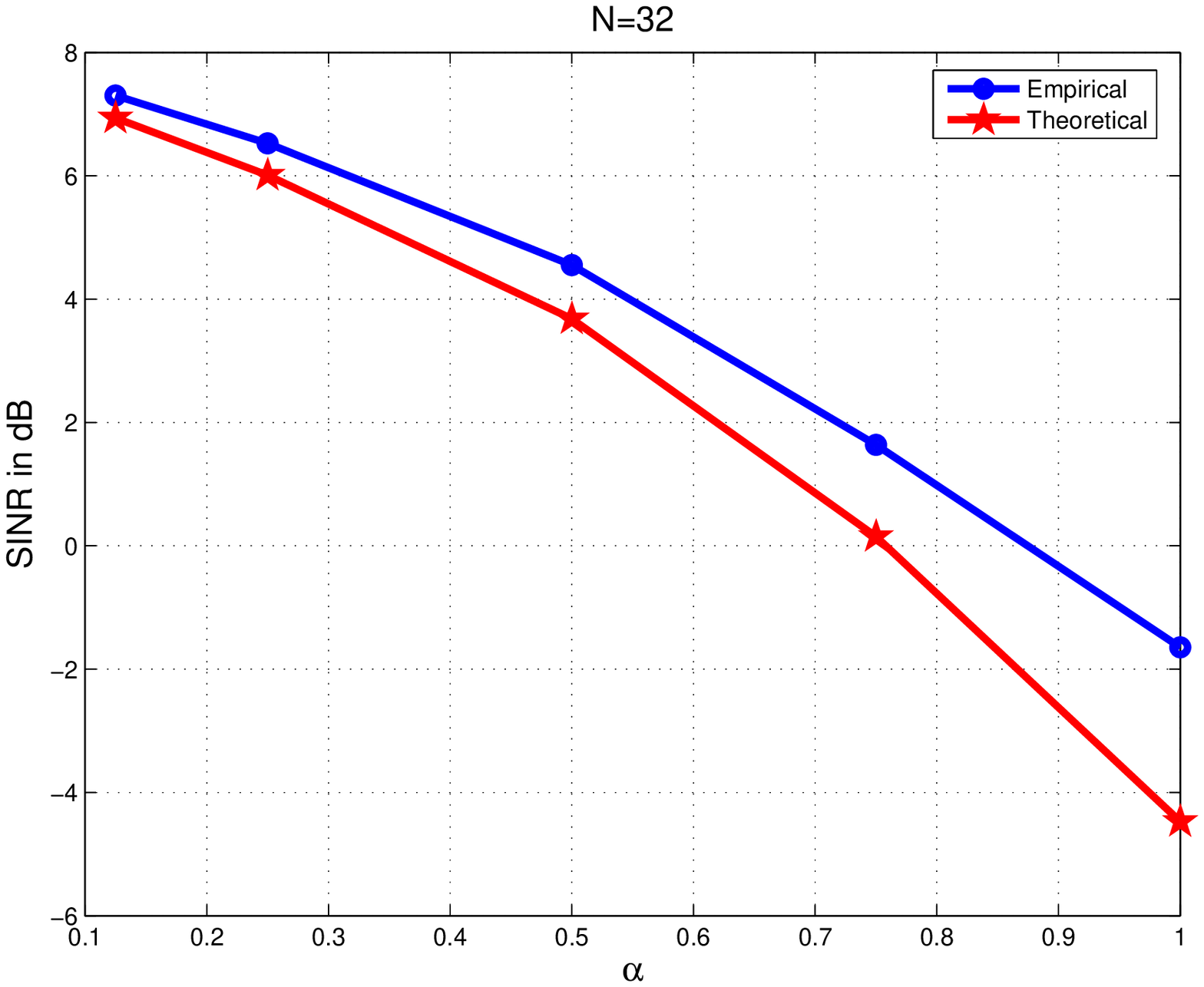}\\
\includegraphics[width=0.4\textwidth]{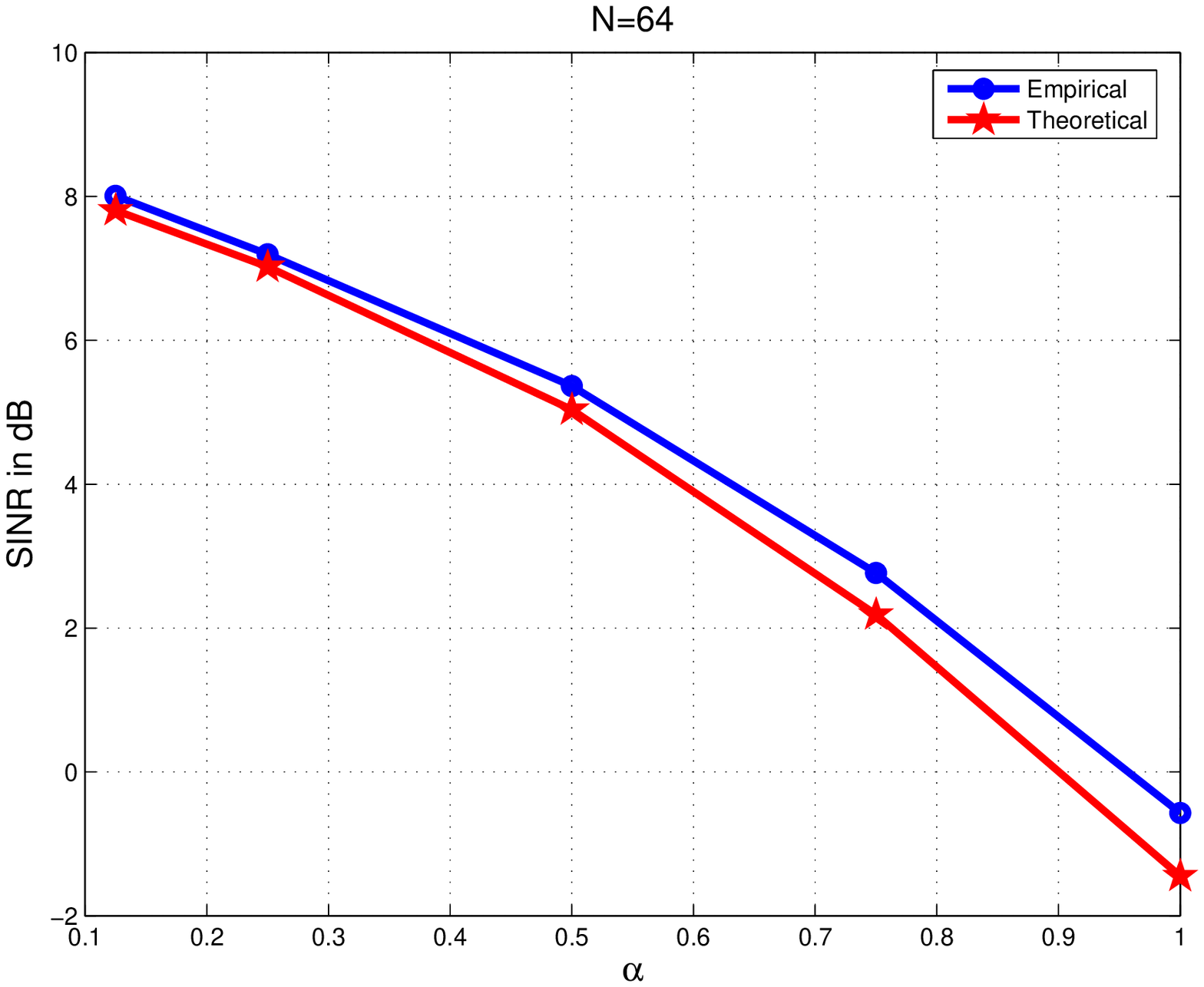}
&
\includegraphics[width=0.4\textwidth]{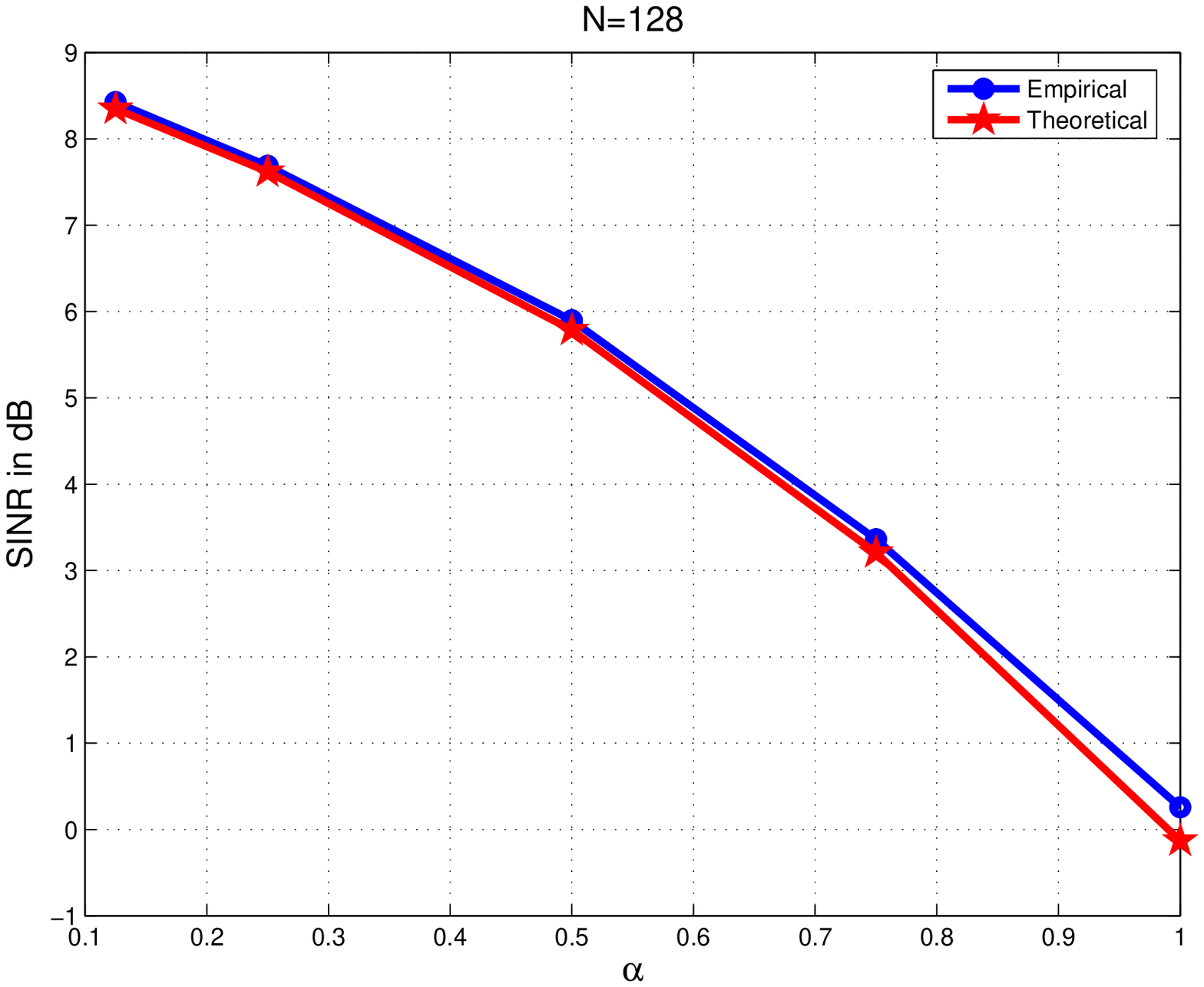}
\end{tabular}
\end{center}
\caption{Theoretical and empirical $1$\% outage SINR}
\label{fig:outage}
\end{figure*}

In order to test the results of Proposition \ref{prop-beta-searable} and
Corollary \ref{cor-clt-separable}, we consider the following multiple antenna
(MIMO) model with exponentially decaying correlation at reception: 
$$
{\bs \Sigma }=\frac{1}{\sqrt{K}}{\bs \Psi}^{1/2} {\bf W} {\bf P}^{1/2} 
$$
where 
${\bs \Psi} = \left[ a^{m-n} \right]_{m,n =0}^{N-1}$ with $0<a<1$ is
the covariance matrix that accounts for the correlations at the receiver side, 
$P=\mbox{diag}\left(p_0,\cdots,p_K\right)$ is the matrix of 
the powers given to the different sources and $W$ is a $N \times (K+1)$ matrix
with Gaussian standard iid elements. 
 Let $\bf {P_u}$ denote the vector containing the powers of the interfering 
 sources. We set $\bf {P_u}$ (up to a permutation of its elements) to:
$$\bf{P_u}=\left\{
\begin{array}{l}
\left[4P \ \  5P\right]  \ \ \ \mathrm{if}  \ \ K=2 \\
\left[P\ \  P\ \  2P\ \  4P\right]  \ \ \ \mathrm{if} \ \ K=4 \\
\left[ P\ \ P \ \ 2P\ \ 2P \ \ 2P\ \  4P\ \ 4P\ \ 4P\ \ 8P \ \  16P \ \ 
16P\ \ 16P \ \right] \  \mathrm{if} \ K=12 \ . \\
\end{array}
\right.
$$
For $K=2^p$ with $3 \le p \le 7$, we assume that the powers of the
interfering sources are arranged into $5$ classes as in Table \ref{tab-powers}.
We set the SNR $P / \rho$ to $10$ dB and $a$ to $0.1$. 
We investigate in this section the accuracy of the Gaussian approximation 
in terms of the outage probability. In Fig.\ref{fig:outage}, we compare 
the empirical $1\%$ outage SINR with the one predicted by the Central Limit 
Theorem. We note that the Gaussian approximation tends to under estimate 
the $1$\% outage SINR. We also note that it has a good accuracy for 
small values of $\alpha$ and for enough large values of $N$ ($N \geq 64$). \\

Observe that all these simulations confirm a fact announced in Remark
\ref{rem-decroissance} above: 
compared with functionals of the channel singular values such as Shannon's 
mutual information, larger signal dimensions are needed to attain the 
asymptotic regime for quadratic forms such as the SINR 
(see for instance outage probability approximations for mutual information in
\cite{mou-sim-sen-it03} and in \cite{mou-sim-it07}). 
This observation holds for first order as well as for second order results. 

\section{Proof of Theorem \ref{th-clt}}
\label{sec-proof}
This section is devoted to the proof of Theorem \ref{th-clt}. We begin with 
mathematical preliminaries. 
\subsection{Preliminaries} 
\label{subsec-preliminaires}
The following lemma gathers useful matrix results, 
whose proofs can be found in \cite{hor-joh-livre94}:  
\begin{lemma}
\label{lm-matrix-analysis} 
Assume ${\bf X} = [ x_{ij} ]_{i,j=1}^N$ and ${\bf Y}$ are complex $N \times N$ 
matrices. Then 
\begin{enumerate}
\item 
\label{lm-matrix-spectral-norm} 
For every $i,j\le N$, $| x_{ij} | \le \| {\bf X} \|$. In particular, 
$\| \diag({\bf X}) \| \le \| {\bf X} \|$. 
\item 
\label{ineq-normes} 
$\| {\bf XY} \| \le \| {\bf X} \| \, \| {\bf Y} \|$. 
\item For $\rho > 0$, the resolvent $({\bf XX}^* + \rho {\bf I})^{-1}$
satisfies 
$\| ({\bf XX}^* + \rho {\bf I})^{-1} \| \le \rho^{-1}$. 
\item 
\label{trXY} 
If ${\bf Y}$ is Hermitian nonnegative, then 
$| \tr({\bf XY}) | \le \| {\bf X} \| \tr({\bf Y})$. 
\end{enumerate} 
\end{lemma} 
\vspace*{0.05\columnwidth} 

Let ${\bf X} = {\bf U} {\bs \Lambda} {\bf V}^*$ be a spectral decomposition of 
${\bf X}$ where ${\bs \Lambda} = \diag(\lambda_1, \ldots, \lambda_n)$ is
the matrix of singular values of ${\bf X}$. For a real $p \geq 1$, the 
Schatten $\ell_p$-norm of ${\bf X}$ is defined as
$\| {\bf X} \|_p = \left( \sum \lambda_i^p \right)^{1/p}$.  
The following bound over the Schatten $\ell_p$-norm of a triangular
matrix will be of help (for a proof, see \cite{bha-gir-kok-spl07},
\cite[page 278]{nik-livre02}):
\begin{lemma}
\label{lm-schatten} 
Let ${\bf X} = [ x_{ij} ]_{i,j=1}^N$ be a $N \times N$ complex matrix
and let $\widetilde{\bf X} = [ x_{ij} {\bf 1}_{i>j} ]_{i,j=1}^N$ be the
strictly lower triangular matrix extracted from ${\bf X}$. Then for every
$p \geq 1$, there exists a constant $C_p$ depending on $p$ only such that
$$
\| \widetilde{\bf X} \|_p \le C_p \| {\bf X} \|_p \ . 
$$
\end{lemma} 

\vspace*{0.05\columnwidth} 
The following lemma lists some properties of the resolvent ${\bf Q}$ and the deterministic 
approximation matrix ${\bf T}$.
Its proof is postponed to Appendix \ref{anx-proof-lm-props-generales-Q-T}. 
\begin{lemma}
\label{lm-props-generales-Q-T}
The following facts hold true: 
\begin{enumerate}
\item 
\label{lm-props-generales-T}
Assume {\bf A2}. Consider matrices ${\bf T}_K(-\rho) = \diag( t_1(-\rho), 
\ldots, t_N(-\rho))$ defined by Theorem 
\ref{th-T-Q}--(\ref{th-1st-order-system}). Then for every $1 \leq n \le N$, 
\begin{equation}
\label{eq-bornes-T} 
\frac{1}{\rho + \sigma_{\max}^2} \le t_n(-\rho) \le \frac{1}{\rho} \ . 
\end{equation} 
\item 
\label{lm-borne-trSQT}
Assume in addition {\bf A1} and {\bf A3}. Let ${\bf Q}_K(-\rho) = 
({\bf YY}^* + \rho {\bf I})^{-1}$ and let matrices ${\bf S}_K$ 
be as in the statement of Theorem 
\ref{th-T-Q}--(\ref{th-1st-order-convergence}). Then 
\begin{equation}
\label{eq-borne-trSQT} 
\sup_K \EE  \left| \tr\ {\bf S}_K 
( {\bf Q}_K - {\bf T}_K) \right| ^2  < \infty \ . 
\end{equation} 
\end{enumerate} 
\end{lemma} 
%
\subsection{Proof of Theorem \ref{th-clt}--(\ref{th-clt-bornes})} 
\label{sec-proof-th-clt-bornes} 
 
We introduce the following notations. 
Assume that ${\bf X}$ is a real matrix, by ${\bf X} \succcurlyeq {\bf 0}$ we mean
$X_{ij} \geq 0$ for every element $X_{ij}$. For a vector ${\bf x}$, 
${\bf x} \succcurlyeq {\bf 0}$ is defined similarly. In the remainder of the
paper, $C = C(\rho, \sigma_{\max}^2, \liminf \frac NK, \sup \frac NK) < \infty$ 
denotes a positive constant 
whose value may change from line to line.

The following lemma, which directly follows from \cite[Lemma 5.2 and
Proposition 5.5]{hac-lou-naj-(clt)-(sub)aap07}, states some important
properties of the matrices ${\bf A}_K$ defined in the statement of
Theorem \ref{th-clt}. 
\begin{lemma}
\label{lm-props-A} 
Assume {\bf A2} and {\bf A3}. Consider matrices ${\bf A}_K$ defined by 
\eqref{eq-def-A}. Then the following facts hold true: 
\begin{enumerate}
\item 
\label{lm-props-A-inversible} 
Matrix ${\bf I}_K - {\bf A}_K$ is invertible, and 
$ ({\bf I}_K - {\bf A}_K)^{-1} \succcurlyeq {\bf 0}$.
\item
\label{lm-props-A-diagonale}
Element $(k,k)$ of the inverse satisfies 
$ \left[ ({\bf I}_K - {\bf A}_K)^{-1} \right]_{k,k} \geq 1$ 
for every $1 \leq k \leq K$. 
\item
\label{lm-props-A-norme-bornee} 
The maximum row sum norm of the inverse satisfies 
$\limsup_K \leftrownorm ({\bf I}_K - {\bf A}_K)^{-1}  \rightrownorm_\infty 
< \infty$. 
\end{enumerate} 
\end{lemma}
\vspace*{0.05\columnwidth} 

Due to Lemma \ref{lm-props-A}--(\ref{lm-props-A-inversible}), $\Theta_K^2$
is well defined. Let us prove that $\limsup_K \Theta_K^2 < \infty$. 
The first term of the right-hand side of \eqref{eq-theta} satisfies 
\begin{multline} 
\frac{1}{K} {\bf g}^\T ( {\bf I}_K - {\bf A}_K)^{-1} {\bs \Delta}^{-1} {\bf g}
\leq 
\| {\bf g} \|_\infty 
\| ( {\bf I}_K - {\bf A}_K)^{-1} {\bs \Delta}^{-1} {\bf g} \|_\infty \\ 
\leq 
\| {\bf g} \|_\infty 
\leftrownorm ({\bf I}_K - {\bf A}_K)^{-1}  \rightrownorm_\infty 
\| {\bs \Delta}^{-1} {\bf g} \|_\infty
\leq
\| {\bf g} \|_\infty^2
\leftrownorm ({\bf I}_K - {\bf A}_K)^{-1}  \rightrownorm_\infty 
\label{eq-inegalites-gAg} 
\end{multline}
due to $\leftrownorm {\bs \Delta}^{-1} \rightrownorm_\infty \le 1$. 
Recall that $\| {\bf T} \| \leq \rho^{-1}$ by Lemma 
\ref{lm-props-generales-Q-T}--(\ref{lm-props-generales-T}). 
Therefore, any element of ${\bf g}$ satisfies
\begin{equation}
\label{eq-limsup-trDDT2} 
\frac 1K \tr {\bf D}_0 {\bf D}_k {\bf T}^2 \leq
\frac NK \| {\bf D}_0 \| \| {\bf D}_k \| \| {\bf T} \|^2 
\leq 
\frac NK \frac{\sigma_{\max}^4}{\rho^2} 
\end{equation} 
by {\bf A2}, hence $\sup_K \| {\bf g} \| \le C$. From Lemma 
\ref{lm-props-A}--(\ref{lm-props-A-norme-bornee}) and \eqref{eq-inegalites-gAg}, 
we then obtain 
\begin{equation}
\label{eq-limsup-gAg} 
\limsup_K 
\frac{1}{K} {\bf g}^\T ( {\bf I}_K - {\bf A}_K)^{-1} {\bs \Delta}^{-1} {\bf g}
\le C .
\end{equation}
We can prove similarly that the second term in the right-hand
side of \eqref{eq-theta} satisfies $\sup_K (( \EE| W_{10} |^4 - 1 )
\frac 1K \tr {\bf D}_0^2 {\bf T}(-\rho)^2 ) \le C$. Hence
$\limsup_K \Theta_K^2 < \infty$. \\
Let us prove that $\liminf_K \Theta_K^2 > 0$. We have
\begin{eqnarray*}
\frac{1}{K} {\bf g}^\T ( {\bf I}_K - {\bf A}_K)^{-1} {\bs \Delta}^{-1} {\bf g}
&\stackrel{(a)}{\geq}& 
\frac{1}{K} {\bf g}^\T 
\diag\left( ( {\bf I}_K - {\bf A}_K)^{-1} \right) {\bs \Delta}^{-1} {\bf g} \\
&\stackrel{(b)}{\geq}& 
\frac{1}{\left(1 + \frac NK \frac{\sigma_{\max}^2}{\rho} \right)^2}
\frac 1K \sum_{k=1}^K 
\left( \frac 1K \tr {\bf D}_0 {\bf D}_k {\bf T}^2 \right)^2  \\
&\stackrel{(c)}{\geq}& 
\frac{1}{\left(1 + \frac NK \frac{\sigma_{\max}^2}{\rho} \right)^2}
\left( \frac{1}{K^2}  
\tr \ {\bf D}_0  
\left(\sum_{k=1}^K {\bf D}_k \right) {\bf T}^2  \right)^2  \\
&\stackrel{(d)}{\geq}& 
\frac{1}{\left(1 + \frac NK \frac{\sigma_{\max}^2}{\rho} \right)^2
(\rho + \sigma^2_{\max})^4}
\left( \frac{1}{K^2}  
\tr \ {\bf D}_0  \sum_{k=1}^K {\bf D}_k  \right)^2  \\
&\geq& {C} 
\left( \frac{1}{K^2}  
\tr \  {\bf D}_0  
\sum_{k=1}^K {\bf D}_k \right)^2\ ,  
\end{eqnarray*} 
where $(a)$ follows from the fact that 
$ ({\bf I}_K - {\bf A}_K)^{-1} \succcurlyeq {\bf 0}$ (Lemma 
\ref{lm-props-A}--(\ref{lm-props-A-inversible}), and the straightforward inequalities
${\bs \Delta}^{-1} \succcurlyeq {\bf 0}$ and ${\bf g} \succcurlyeq {\bf 0}$), 
$(b)$ follows from Lemma \ref{lm-props-A}--(\ref{lm-props-A-diagonale}) and  
$\| {\bs \Delta} \| \leq (1 + \frac NK \frac{\sigma_{\max}^2}{\rho} )^2$, 
$(c)$ follows from the elementary inequality 
$n^{-1} \sum x_i^2 \geq (n^{-1} \sum x_i )^2$, and $(d)$ is due to Lemma
\ref{lm-props-generales-Q-T}--(\ref{lm-props-generales-T}). 
Similar derivations yield:
$$
( \EE| W_{10} |^4 - 1 )
\frac 1K \tr {\bf D}_0^2 {\bf T}\quad \geq  \quad
\frac{ \EE| W_{10} |^4 - 1 }{(\rho+\sigma_{\max}^2)^2} 
\left( \frac 1K \tr {\bf D}_0 \right)^2 
\quad \geq \quad C ( \EE| W_{10} |^4 - 1 )
$$ 
by {\bf A3}. Therefore, if {\bf A4} holds true, then $\liminf_K \Theta_K^2 > 0$ and
Theorem \ref{th-clt}--(\ref{th-clt-bornes}) is proved.

\subsection{Proof of Theorem \ref{th-clt}--(\ref{th-clt-cvg})} 
\label{sec-proof-th-clt-cvg} 

Recall that the SINR $\beta_K$ is given by Equation \eqref{eq-fq}. 
The random variable $\frac{\sqrt{K}}{\Theta_K}
(\beta_K - \overline\beta_K)$ can therefore be decomposed as
\begin{eqnarray} 
\frac{\sqrt{K}}{\Theta_K}(\beta_K - \overline\beta_K) &=& 
\frac{1}{\sqrt{K} \Theta_K} 
\left( 
{\bf w}_{0}^* {\bf D}_{0}^{1/2} {\bf Q} {\bf D}_{0}^{1/2} {\bf w}_{0} 
- 
\tr({\bf D}_{0} {\bf Q}) 
\right) 
+ 
\frac{1}{\sqrt{K} \Theta_K} 
\left( 
\tr({\bf D}_{0} ( {\bf Q} - {\bf T}) ) \right) \nonumber \\
&=& 
U_{1,K} + U_{2,K} \label{decompo} \ . 
\end{eqnarray} 
Thanks to Lemma \ref{lm-props-generales-Q-T}--(\ref{lm-borne-trSQT}) and to
the fact that $\liminf_K \Theta_K^2 > 0$, we have 
$\EE U_{K,2}^2 < C K^{-1}$ which implies that $U_{K,2}\rightarrow 0$ in probability as ${K\rightarrow \infty}$. 
Hence, in order to conclude that 
$$
\frac{\sqrt{K}}{\Theta_K}(\beta_K - \overline\beta_K) 
\xrightarrow[K\to\infty]{} {\mathcal N}(0,1) \quad \text{in distribution}\ ,
$$
it is sufficient by Slutsky's theorem to prove that $U_{1,K} \to {\cal
  N}(0,1)$ in distribution. The remainder of the section is devoted to this point.

\begin{remark}
  Decomposition \eqref{decompo} and the convergence to zero (in
  probability) of $U_{2,K}$ yield the following interpretation: The
  fluctuations of $\sqrt{K}(\beta_K - \overline\beta_K)$ are mainly
  due to the fluctuations of vector ${\bf w}_0$. Indeed the
  contribution of the fluctuations\footnote{In fact, one may prove
    that the fluctuation of $\frac 1K \tr {\bf D}_0 ({\bf Q}- {\bf
      T})$ are of order $K$, i.e. $\tr {\bf D}_0 ({\bf Q}- {\bf
      T})$ asymptotically behaves as a Gaussian random variable.  Such
    a speed of fluctuations already appears in
    \cite{hac-lou-naj-(clt)-(sub)aap07}, when studying the
    fluctuations of the mutual information.} of $\frac 1K \tr {\bf
    D}_0 {\bf Q}$, due to the random nature of ${\bf Y}$, is
  negligible.
\end{remark} 
\vspace*{0.05\columnwidth} 
Denote by $\EE_n$ the conditional expectation 
$\EE_n[\ \cdot\ ] = \EE[\ \cdot\ \|\ W_{n,0}, W_{n+1,0}, \ldots, W_{N,0}, {\bf Y} ]$. 
Put $\EE_{N+1}[\ \cdot\ ] = \EE[\ \cdot \ \|\ {\bf Y}]$ and note that
$\EE_{N+1} ( {\bf w}_0^* {\bf D}_0^{1/2} {\bf Q} {\bf D}_0^{1/2} {\bf w}_0 ) =
\tr {\bf D}_0 {\bf Q}$. With these notations at hand, we have:
\begin{equation}
\label{eq-def-Z} 
U_{1,K} = 
\frac{1}{\Theta_K}
\sum_{n=1}^N (\EE_n - \EE_{n+1})
\frac{{\bf w}_0^* {\bf D}_0^{1/2} {\bf Q} {\bf D}_0^{1/2} {\bf w}_0}{\sqrt{K}}
\stackrel{\triangle}{=} 
\frac{1}{\Theta_K} \sum_{n=1}^N Z_{n,K} \ .
\end{equation} 

Consider the increasing sequence of $\sigma-$fields 
$$
{\cal F}_{N,K} =
\sigma(W_{N,0}, {\bf Y})\ ,\quad  \cdots\ ,\quad {\cal F}_{1,K} =
\sigma(W_{1,0} ,\cdots, W_{N,0}, {\bf Y})\ .
$$ Then the random variable
$Z_{n,K}$ is integrable and measurable with respect to ${\cal
  F}_{n,K}$; moreover it readily satisfies $\EE_{n+1} Z_{n,K} = 0$.
In particular, the sequence $(Z_{N,K}, \ldots, Z_{1,K})$ is a martingale 
difference sequence with respect to $\left( {\cal F}_{N,K}, \cdots, {\cal F}_{1,K}\right)$.
The following CLT for martingales is the key tool to study
the asymptotic behavior of $U_{1,K}$: 
\begin{theorem}
\label{th-clt-martingales}
Let $X_{N,K}, X_{N-1,K}, \ldots, X_{1,K}$ be a
martingale difference sequence with respect to the increasing filtration
${\mathcal G}_{N,K}, \ldots, {\mathcal G}_{1,K}$.
Assume that there exists a sequence of real positive numbers
$s_K^2$ such that
$$ 
\frac{1}{s_K^2}
\sum_{n=1}^N \EE \left[ X_{n,K}^2 \| {\mathcal G}_{n+1,K} \right]  
\xrightarrow[K\to\infty]{}  1 
$$ 
in probability. Assume further that the Lyapunov condition holds:
$$ 
\exists \alpha > 0, \quad
\frac{1}{s_K^{2(1+\alpha)}}
\sum_{n=1}^N \EE\left| X_{n,K} \right|^{2+\alpha}
\xrightarrow[K\rightarrow \infty]{} 0\ ,
$$ 
Then $s_K^{-1} \sum_{n=1}^N X_{n,K}$ converges in distribution to
${\mathcal N}(0,1)$ as $K \to \infty$. 
\end{theorem}
\vspace*{0.05\columnwidth} 
\begin{remark}
This theorem is proved in
\cite{bil-PM-livre95}, gathering Theorem 35.12 (which is expressed
under the weaker Lindeberg condition) together with the arguments of Section 27
(where it is proved that Lyapunov's condition implies Lindeberg's condition).
\end{remark}

\vspace*{0.05\columnwidth} 
In order to prove that
\begin{equation}
\label{eq-cvg-U1->N(0,1)} 
U_{1,K} = 
\frac{1}{\Theta_K} \sum_{n=1}^N Z_{n,K} 
\xrightarrow[K\to\infty]{} {\cal N}(0,1) 
\quad \text{in distribution}  \ ,
\end{equation}
we shall apply Theorem \ref{th-clt-martingales} to the sum 
$\frac{1}{\Theta_K} \sum_{n=1}^N Z_{n,K}$ and the filtration $\left( {\cal F}_{n,K} \right)$. 
The proof is carried out into four steps:

\paragraph*{Step 1} We first establish Lyapunov's condition. Due to the fact that
$\liminf_K \Theta_K^2 > 0$, we only need to show that
\begin{equation}
\label{eq-lyapunov-Z} 
\exists\ \alpha > 0,\quad  \sum_{n=1}^N \EE | Z_{n,K} |^{2+\alpha} 
\xrightarrow[K\to\infty]{} 0\ . 
\end{equation}

\paragraph*{Step 2}  
We prove that $V_K = \sum_{n=1}^N \EE_{n+1} Z_{n,K}^2$ satisfies 
\begin{equation}
\label{eq-sumZ-trDQDQ}
V_K - 
\left( 
\frac{\left( \EE | W_{10} |^4 - 2 \right)}{K}  
\tr \left( {\bf D}_0^2 (\diag({\bf Q}))^2 \right) 
+ 
\frac{1}{K} \tr ({\bf D}_0 {\bf QD}_0 {\bf Q}) \right) 
\xrightarrow[K\to\infty]{} 0 
\quad \text{in probability} \ . 
\end{equation} 

\paragraph*{Step 3} We first show that 
\begin{equation}
\label{eq-D2Q2-D2T2} 
\frac{1}{K} \tr {\bf D}_0^2 (\diag({\bf Q}))^2 - 
\frac{1}{K} \tr {\bf D}_0^2 {\bf T}^2 
\xrightarrow[K\to\infty]{} 0 \quad \text{in probability.} 
\end{equation} 
In order to study the asymptotic behavior of 
$\frac{1}{K} \tr ({\bf D}_0 {\bf QD}_0 {\bf Q})$, 
we introduce the random variables 
$U_\ell = \frac{1}{K} \tr ({\bf D}_0 {\bf QD}_\ell {\bf Q})$ for 
$0 \le \ell \le K$ (the one of interest being $U_0$). We then prove that the $U_{\ell}$'s satisfy the following 
system of equations:
\begin{equation}
\label{eq-syst-Ul}
U_\ell = \sum_{k=1}^K c_{\ell k} U_k + 
\frac 1K \tr {\bf D}_0 {\bf D}_\ell {\bf T}^2 + \epsilon_\ell ,
\quad 0 \le \ell \le K, 
\end{equation} 
where 
\begin{equation}
\label{eq-clk} 
c_{\ell k} = 
\frac 1K \frac{\frac{1}{K} \tr {\bf D}_{\ell} {\bf D}_{k} {\bf T}(-\rho)^2}
{\left(1 + \frac{1}{K} \tr {\bf D}_{k} {\bf T}(-\rho) \right)^2}, 
\quad 0 \le \ell\le K , \ 1\le k \le K 
\end{equation} 
and the perturbations $\epsilon_\ell$ satisfy $\EE | \epsilon_\ell | 
\le CK^{-\frac 12}$ where we recall that $C$ is independent of $\ell$. 

\paragraph*{Step 4} 
We prove that $U_0 = \frac 1K \tr {\bf D}_0 {\bf Q} {\bf D}_0 {\bf Q}$ 
satisfies 
\begin{equation}
\label{eq-U0} 
U_0 = 
\frac 1K \tr {\bf D}_0^2 {\bf T}^2 + \frac 1K {\bf g}^\T 
\left({\bf I} - {\bf A}\right)^{-1} {\bs \Delta}^{-1} {\bf g} 
+ \epsilon 
\end{equation} 
with $\EE|\epsilon| \le C K^{-\frac 12}$. This equation combined with  
\eqref{eq-sumZ-trDQDQ} and \eqref{eq-D2Q2-D2T2} yields
$\sum_n \EE_{n+1} Z_{n,K}^2 - \Theta_K^2 \to 0$ in probability. 
As $\liminf_K \Theta_K^2 > 0$, this implies 
$\frac{1}{\Theta_K} \sum_n \EE_{n+1} Z_{n,K}^2 \to 1$ in probability, 
which proves \eqref{eq-cvg-U1->N(0,1)} and thus ends the proof of
Theorem \ref{th-clt}. 

\vspace*{0.05\columnwidth} 

Write ${\bf B} = [ b_{ij} ]_{i,j=1}^N = {\bf D}_0^{1/2} {\bf Q} 
{\bf D}_0^{1/2}$ and recall from \eqref{eq-def-Z} that 
$Z_{n,K} = \frac{1}{\sqrt{K}} (\EE_n - \EE_{n+1}) 
{\bf w}_0^* {\bf B} {\bf w}_0$. We have 
$$ 
\EE_n {\bf w}_0^* {\bf B} {\bf w}_0 
=
\sum_{\ell = 1}^{n-1} b_{\ell \ell} 
+ 
\sum_{\ell_1, \ell_2 = n}^N 
 W_{\ell_1 0}^* W_{\ell_2 0} b_{\ell_1 \ell_2} \ .
$$ 
Hence 
\begin{equation}
\label{eq-expression-Zn} 
Z_{n,K} = \frac{1}{\sqrt{K}} \left( 
\left( | W_{n0} |^2 - 1 \right) b_{nn} + 
W_{n0}^* \sum_{\ell=n+1}^N W_{\ell 0} b_{n \ell} + 
W_{n0} \sum_{\ell=n+1}^N W_{\ell 0}^* b_{\ell n} \right) \ .
\end{equation} 

\paragraph*{Step 1: Validation of the Lyapunov condition} The
following inequality will be of help to check Lyapunov's condition.
\begin{lemma}[Burkholder's inequality]
\label{lm-burkholder} 
Let $X_k$ be a complex martingale difference sequence with respect to the 
increasing sequence of $\sigma$--fields ${\cal F}_k$. Then for $p\ge 2$, 
there exists a constant $C_p$ for which 
$$
\EE\left| \sum_k X_k \right|^p 
\le 
C_p 
\left(
\EE\left( \sum_k \EE\left[ | X_k|^2 \| {\cal F}_{k-1} \right] \right)^{p/2}
+ 
\EE \sum_k | X_k|^p 
\right) \ . 
$$
\end{lemma}  
\vspace*{0.05\columnwidth} 
Recall  Assumption {\bf A1}. Eq. \eqref{eq-expression-Zn} yields:
\begin{eqnarray}
\left| Z_{n,K} \right|^{4}
&\leq& 
\frac{1}{K^2} 
\left( 
\frac{| W_{n0} |^2 + 1}{\rho \sigma_{\max}^2} +
2 
\left| W_{n0}  
\sum_{\ell=n+1}^N W_{\ell 0} b_{n \ell} \right| 
\right)^4 \nonumber \\
&\leq& 
\frac{2^3}{K^2} 
\left( 
\left( \frac{| W_{n0} |^2 + 1}{\rho \sigma_{\max}^2} \right)^4
+
2^4
\left| W_{n0} \sum_{\ell=n+1}^N W_{\ell 0} b_{n \ell} \right|^4
\right) 
\label{eq-borne-Z-nK} 
\end{eqnarray} 
where we use the fact that $|b_{nn}|\le (\rho \sigma^2_{\max})^{-1}$
(cf. Lemma \ref{lm-matrix-analysis}--(\ref{lm-matrix-spectral-norm}))
and the convexity of $x \mapsto x^4$. Due to Assumption
{\bf A1}, we have:
\begin{equation}
\label{eq-borne-terme1} 
\EE 
\left( | W_{n0} |^2 + 1 \right)^4 \leq
2^3 \left( \EE | W_{n0} |^8 + 1 \right)
<  \infty \ .
\end{equation} 
Considering the second term at the right-hand side of 
\eqref{eq-borne-Z-nK}, we write 
\begin{eqnarray*}
\EE 
\left| W_{n0} \sum_{\ell=n+1}^N W_{\ell 0} b_{n \ell} 
\right|^4 &=&
\EE \left| W_{n0} \right|^4
\EE \left| \sum_{\ell=n+1}^N W_{\ell 0} b_{n \ell} 
\right|^4 \ ,\\
&\stackrel{(a)}{\leq}& 
C \left( \EE \left( \sum_{\ell=n+1}^N (\EE| W_{\ell 0}|^2)  
|b_{n \ell} |^2 \right)^2 + 
\sum_{\ell=n+1}^N (\EE| W_{\ell 0}|^4) 
(\EE |b_{n \ell} |^4)  
\right) \ ,\\ 
&\stackrel{(b)}{\leq}& 
C \left( \EE \left( \sum_{\ell=n+1}^N   
|b_{n \ell} |^2 \right)^2 + 
\sum_{\ell=n+1}^N  
\EE |b_{n \ell} |^{2} \right)  \ ,
\end{eqnarray*}
where $(a)$ follows from Lemma \ref{lm-burkholder} (Burkholder's inequality), the filtration being 
${\cal F}_{N,K}, \ldots,$ ${\cal F}_{n+1,K}$ and $(b)$ follows from the bound 
$|b_{n \ell} |^4
\le |b_{n \ell} |^{2} \max\,|b_{n \ell}|^2
\le |b_{n \ell} |^{2} (\sigma^2_{\max} \rho^{-1})^2$
(cf. Lemma \ref{lm-matrix-analysis}--(\ref{lm-matrix-spectral-norm})).  
Now, notice that
$$
\sum_{\ell=n+1}^N  |b_{n \ell} |^{2} 
< 
\sum_{\ell=1}^N  |b_{n \ell} |^{2} 
= 
 \left[ {\bf D}_0^{1/2} {\bf Q} {\bf D}_0 {\bf Q} {\bf D}_0^{1/2} 
\right]_{nn} \le 
 \| {\bf D}_0^{1/2} {\bf Q} {\bf D}_0 {\bf Q} {\bf D}_0^{1/2} \|
\le
\frac{\sigma_{\max}^4}{\rho^2} \ . 
$$
This yields 
$\EE | W_{n0} \sum_{\ell=n+1}^N W_{\ell\, 0} b_{n \ell} 
|^4 \le C$. Gathering this result with \eqref{eq-borne-terme1}, getting back to
\eqref{eq-borne-Z-nK}, taking the expectation and summing up finally yields:
$$
\sum_{n=1}^N \EE |Z_{n,K}|^4
\ \le \ \frac{C}{K} \ \xrightarrow[K\to\infty]{} \ 0 
$$
which establishes Lyapunov's condition \eqref{eq-lyapunov-Z} with
$\alpha = 2$.

\paragraph*{Step 2: Proof of \eqref{eq-sumZ-trDQDQ}} 
Eq. \eqref{eq-expression-Zn} yields:
\begin{eqnarray*} 
\EE_{n+1} Z_{n,K}^2 &=&  
\frac 1K \left( \phantom{\sum_{\ell=n+1}^N} 
\! \! \!  \! \! \!  \! \! \!  \! \! \! 
\! \! \!  \! \!   
\right. 
\left( \EE | W_{10} |^4 - 1 \right) b_{nn}^2 
+ 
\EE_{n+1} 
\left( W_{n0}^* \sum_{\ell=n+1}^N W_{\ell 0} b_{n \ell} + 
W_{n0} \sum_{\ell=n+1}^N W_{\ell 0}^* b_{\ell n} \right)^2
\\
& & + 
2 b_{nn} \left( \EE\, W_{10}^* | W_{10} |^2 \right) 
\sum_{\ell=n+1}^N W_{\ell 0} b_{n \ell}
+ 
2 b_{nn} \left( \EE\, W_{10} | W_{10} |^2 \right) 
\sum_{\ell=n+1}^N W_{\ell 0}^* b_{\ell n} 
\left. \phantom{\sum_{\ell=n+1}^N} 
\! \! \!  \! \! \!  \! \! \!  \! \! \! 
\! \! \!  \! \! \!  
\right) \ .
\end{eqnarray*} 
Note that the second term of the right-hand side writes:
$$
\EE_{n+1} 
\left( W_{n0}^* \sum_{\ell=n+1}^N W_{\ell 0} b_{n \ell} + 
W_{n0} \sum_{\ell=n+1}^N W_{\ell 0}^* b_{\ell n} \right)^2
= 
2 \sum_{\ell_1, \ell_2 = n+1}^N W_{\ell_1 0} W_{\ell_2 0}^* 
b_{n \ell_1} b_{\ell_2 n} \ .
$$
Therefore, $V_K = \sum_{n=1}^N \EE_{n+1} Z_{n,K}^2$ writes:
\begin{multline*} 
V_K = 
\frac{\left( \EE | W_{10} |^4 - 1 \right)}{K}  \sum_{n=1}^N b_{nn}^2 
+ \frac{2}{K} 
\sum_{n=1}^N \sum_{\ell_1, \ell_2 = n+1}^N W_{\ell_1 0} W_{\ell_2 0}^* 
b_{n \ell_1} b_{\ell_2 n} \\ 
+ \frac{2}{K}
\Re \left( 
\left( \EE\,W_{10}^* | W_{10} |^2 \right) 
\sum_{n=1}^N b_{nn} 
\sum_{\ell=n+1}^N W_{\ell 0} b_{n \ell}
\right) \ ,
\end{multline*} 
where $\Re$ denotes the real part of a complex number.
We introduce the following notations:
$$
{\bf R} = \left( r_{ij} \right)_{i,j=1}^N \stackrel{\triangle}= \left( b_{ij} {\bf 1}_{i > j} \right)_{i,j=1}^N
\qquad \textrm{and}\qquad 
\Gamma_K = \frac{1}{K}
\sum_{n=1}^N b_{nn} 
\sum_{\ell=n+1}^N W_{\ell 0} b_{n \ell}\ .
$$
Note in particular that ${\bf R}$ is the strictly lower triangular
matrix extracted from ${\bf D}_0^{1/2} {\bf Q} {\bf D}_0^{1/2}$. We
can now rewrite $V_K$ as:
\begin{equation}
\label{eq-expression-VK} 
V_K = 
\frac{\left( \EE | W_{10} |^4 - 1 \right)}{K}  
\tr \left( {\bf D}_0^2 (\diag({\bf Q}))^2 \right) 
+ \frac{2}{K} {\bf w}_0^* {\bf RR}^* {\bf w}_0 
+ 2 \Re\left( \Gamma_K \EE W_{10}^* | W_{10} |^2 \right) \ .
\end{equation} 
We now prove that the third term of the right-hand side vanishes,
and find an asymptotic equivalent for the second one. Using Lemma \ref{lm-matrix-analysis}, we have:
\begin{eqnarray*} 
\EE_{N+1} | \Gamma_K |^2 &=& 
\frac{1}{K^2} 
\sum_{n, m = 1}^N b_{nn} b_{mm} \sum_{\ell=1}^N 
b_{n \ell} b_{m \ell}^* {\bf 1}_{\ell > n} {\bf 1}_{\ell > m} 
= 
\frac{1}{K^2} \tr \left( \diag({\bf B})  {\bf R}^* {\bf R} \diag({\bf B}) 
\right) \\
&=&
\frac{1}{K^2} \tr 
\left( {\bf D}_0^{1/2} \diag({\bf Q}) {\bf D}_0^{1/2}
{\bf R}^* {\bf R} 
{\bf D}_0^{1/2} \diag({\bf Q}) {\bf D}_0^{1/2}
\right) \\
&\leq& 
\frac{1}{K^2} 
\| {\bf D}_0 \|^2 \| {\bf Q} \|^2 \tr( {\bf R}^* {\bf R} )  
\quad \leq\quad 
\frac{1}{K^2} 
\| {\bf D}_0 \|^2 \| {\bf Q} \|^2 \tr( {\bf B}^2 ) \quad \le \quad 
\frac{1}{K^2} 
\| {\bf D}_0 \|^4 \| {\bf Q} \|^2 \tr( {\bf Q}^2 )\\
&\leq& 
\frac{1}{K} 
\| {\bf D}_0 \|^2 \| {\bf Q} \|^4  
\quad \leq\quad 
\frac{1}{K} 
\frac{\sigma_{\max}^4}{\rho^4} \quad  \xrightarrow[K\to\infty]{}\quad  0\ . 
\end{eqnarray*} 
In particular, $\EE| \Gamma_K |^2\rightarrow 0 $ and 
\begin{equation}
\label{eq-cvg-EK} 
\Re\left( \left( \EE W_{10}^* | W_{10} |^2 \right)  \Gamma_K \right) 
\xrightarrow[K\to\infty]{} 0 \quad \text{in probability} \ . 
\end{equation} 
Consider now the second term of the right-hand side of 
Eq. \eqref{eq-expression-VK}. We prove that:
\begin{equation}
\label{eq-cvg-wRRw} 
\frac{1}{K} {\bf w}_0^* {\bf RR}^* {\bf w}_0 - \frac{1}{K} \tr({\bf RR}^*) 
\xrightarrow[K\to\infty]{} 0 \quad 
\text{in probability.}
\end{equation} 
By Lemma \ref{lm-jack} (Ineq. \eqref{eq-jack-longue}), we have  
$$
\EE\left( 
\frac{1}{K} {\bf w}_0^* {\bf RR}^* {\bf w}_0 - \frac{1}{K} \tr({\bf RR}^*) 
\right)^{2} \le
\frac{C}{K^2} 
(\EE| W_{10} |^4) \tr( {\bf RR}^* {\bf RR}^* )  \ . 
$$
Notice that $\tr( {\bf RR}^* {\bf RR}^*) = \| {\bf R} \|_4^4$ where
$\| {\bf R} \|_4$ is the Schatten $\ell_4$-norm of ${\bf R}$. 
Using Lemma \ref{lm-schatten}, we have: 
$$
\| {\bf R} \|_4^4 \leq C \| {\bf D}_0^{1/2} {\bf Q} {\bf D}_0^{1/2} \|_4^4
\le N C \| {\bf D}_0^{1/2} {\bf Q} {\bf D}_0^{1/2} \|^4 
\le N \frac{C \sigma_{\max}^8}{\rho^4}\ . 
$$
Therefore, 
$$
\EE\left( 
\frac{1}{K} {\bf w}_0^* {\bf RR}^* {\bf w}_0 - \frac{1}{K} \tr({\bf RR}^*) 
\right)^{2} \le
C \frac{N}{K^2} 
\xrightarrow[K\to\infty]{} 0 
$$
which implies \eqref{eq-cvg-wRRw}. Now, due to the fact that 
${\bf B} = {\bf B}^*$, we have 
\begin{eqnarray}
\frac{2}{K} \tr \, {\bf RR}^* &=& 
\frac 2K \sum_{n=1}^{N} \sum_{\ell = n+1}^N | b_{n \ell} |^2 \nonumber \\
&=& 
\frac{1}{K} \sum_{n, \ell=1}^N | b_{n \ell} |^2 - 
\frac 1K \sum_{n=1}^N | b_{nn} |^2 \nonumber \\
&=&
\frac 1K \tr\,{\bf D}_0{\bf QD}_0 {\bf Q} - 
\frac 1K \tr\,{\bf D}_0^2 (\diag({\bf Q}))^2 
\label{eq-trRR} 
\end{eqnarray}
Gathering (\ref{eq-expression-VK}--\ref{eq-trRR}), we obtain 
\eqref{eq-sumZ-trDQDQ}. Step 2 is proved.

\paragraph*{Step 3: Proof of \eqref{eq-D2Q2-D2T2} and \eqref{eq-syst-Ul}} 
We begin with some identities. Write 
${\bf Q}(z) = [ q_{ij}(z) ]_{i,j=1}^N$ and 
$\widetilde{\bf Q}(z) = [ \tilde q_{ij}(z) ]_{i,j=1}^K$. Denote by 
${\bf y}_k$ the column number $k$ of ${\bf Y}$ and by ${\bs \xi}_n$ the
row number $n$ of ${\bf Y}$. 
Denote by ${\bf Y}^{k}$ the matrix that remains after deleting column
$k$ from ${\bf Y}$ and by ${\bf Y}_{n}$ the matrix that remains after 
deleting row $n$ from ${\bf Y}$. 
Finally, write ${\bf Q}_k(z) = ( {\bf Y}^k {{\bf Y}^k}^* - z {\bf I})^{-1}$ and
$\widetilde{\bf Q}_n(z) = ( {\bf Y}_n^* {\bf Y}_n - z {\bf I})^{-1}$. 
The following formulas can be established easily (see for instance 
\cite[\S 0.7.3. and \S 0.7.4]{hor-joh-livre94}):
\begin{equation}
\label{eq-qnn} 
q_{nn}(-\rho) = 
\frac{1}
{\rho( 1 + {\bs \xi}_n \widetilde{\bf Q}_n(-\rho) {\bs \xi}_n^* )}, 
\quad 
\tilde q_{kk}(-\rho) = 
\frac{1}
{\rho( 1 + {\bf y}_k^* {\bf Q}_k(-\rho) {\bf y}_k )}, 
\end{equation} 
\begin{equation}
\label{eq-Q-lemme-inversion} 
{\bf Q} = {\bf Q}_k - 
\frac{{\bf Q}_k {\bf y}_k {\bf y}_k^* {\bf Q}_k}
{1+{\bf y}_k^* {\bf Q}_k {\bf y}_k} 
\end{equation} 
\begin{lemma}
\label{lm-props-Q} 
The following hold true: 
\begin{enumerate}
\item 
\label{rank-one} 
(\emph{Rank one perturbation inequality}) 
The resolvent ${\bf Q}_k(-\rho)$ satisfies 
$\left| \tr {\bf A}( {\bf Q} - {\bf Q}_k ) \right| \le \| {\bf A} \| / \rho$
for any $N \times N$ matrix ${\bf A}$.
\item 
\label{(q-t)-individuels} 
Let Assumptions {\bf A1}--{\bf A3} hold. Then,
\begin{equation}
\label{eq-E(q-t)} 
\max_{1\le n \le N} \EE ( q_{nn}(-\rho) - t_{n}(-\rho) )^2 
\le \frac{C}{K} \ . 
\end{equation} 
The same conclusion holds true if $q_{nn}$ and $t_n$ are replaced with 
$\tilde q_{kk}$ and $\tilde t_k$ respectively. 
\end{enumerate}
\end{lemma} 
\vspace*{0.05\columnwidth} 

We are now in position to prove \eqref{eq-D2Q2-D2T2}. 
First, notice that:
\begin{eqnarray}\label{majo}
\EE \left| q_{nn}^2 -t_n^2 \right| & = & 
\EE \left| q_{nn} -t_n \right| \left( q_{nn} + t_n \right)\nonumber \\
&\le & \sqrt{\EE (q_{nn} -t_n)^2 } \sqrt{\EE (q_{nn} + t_n)^2 } \quad 
\le \quad 
\frac 2{\rho} \sqrt{\EE (q_{nn} -t_n)^2 }\ . 
\end{eqnarray}  
Now, 
\begin{eqnarray*}
\frac{1}{K} \EE 
\left| \tr\, {\bf D}_0^2 ( \diag({\bf Q})^2 - {\bf T}^2 )  \right|
&\le & \frac{1}{K}  \sum_{n=1}^N \sigma_{0,n}^4  
\EE \left| q_{nn}^2 - t_n^2 \right|
\quad \le \quad 
\frac{\sigma_{\max}^4 N}K \max_{1\le n\le N}
\EE \left| q_{nn}^2 - t_n^2 \right| \\
&\le &  \frac{2 \sigma_{\max}^4 N}{\rho K} 
\sqrt{\max_{1\le n\le N}\EE ( q_{nn} - t_n)^2  } \quad
\xrightarrow[K\rightarrow\infty]{}\quad  0\ ,
\end{eqnarray*}
where the last inequality follows from \eqref{majo} together with 
Lemma \ref{lm-props-Q}--(\ref{(q-t)-individuels}).  
Convergence \eqref{eq-D2Q2-D2T2} is established.

We now establish the system of equations \eqref{eq-syst-Ul}. 
Our starting point is the identity
$$
{\bf Q} = {\bf T} + {\bf T} ( {\bf T}^{-1} - {\bf Q}^{-1} ) {\bf Q} 
= 
{\bf T} + \frac{\rho}{K} {\bf T}\, 
\diag( \tr\widetilde{\bf D}_1 \widetilde{\bf T}, \ldots, 
\tr\widetilde{\bf D}_N \widetilde{\bf T} ){\bf Q} - {\bf T}{\bf YY}^* 
{\bf Q} \ . 
$$
Using this identity, we develop $U_\ell = \frac 1K \tr {\bf D}_0 {\bf Q}
{\bf D}_\ell {\bf Q}$ as
\begin{eqnarray}
U_\ell &=& \frac 1K \tr {\bf D}_0 {\bf QD}_\ell {\bf T} + 
\frac{\rho}{K^2}  \tr {\bf D}_0 {\bf QD}_\ell {\bf T} 
\diag( \tr\widetilde{\bf D}_1 \widetilde{\bf T}, \ldots, 
\tr\widetilde{\bf D}_N \widetilde{\bf T} ) {\bf Q} 
- 
\frac 1K \tr {\bf D}_0 {\bf QD}_\ell {\bf T} {\bf YY}^* {\bf Q} 
\nonumber \\
&\stackrel{\triangle}=& 
X_1 + X_2 - X_3 \ . 
\label{eq-Ul} 
\end{eqnarray} 
Lemma \ref{lm-props-generales-Q-T}--(\ref{lm-borne-trSQT}) with
${\bf S} = {\bf D}_0 {\bf D}_\ell {\bf T}$ yields:
\begin{equation}
\label{eq-X1}
X_1 = \frac 1K \tr {\bf D}_0 {\bf D}_\ell {\bf T}^2 + \epsilon_1
\end{equation}
where $\EE | \epsilon_1 | \le \sqrt{\EE \epsilon_1^2} \le C / K$. 
Consider now the term $X_3 = 
\frac 1K \sum_{k=1}^K \tr {\bf D}_0 {\bf QD}_\ell {\bf T} {\bf y}_k
{\bf y}_k^* {\bf Q}$. Using \eqref{eq-qnn} and \eqref{eq-Q-lemme-inversion},
we have 
$$
{\bf y}_k^* {\bf Q} = \left( 1 - 
\frac{{\bf y}_k^* {\bf Q} {\bf y}_k}{1 + {\bf y}_k^* {\bf Q} {\bf y}_k} 
\right) {\bf y}_k^* {\bf Q}_k = \rho\, \tilde{q}_{kk}\, {\bf y}_k^* {\bf Q}_k\ .
$$
Hence 
\begin{eqnarray} 
X_3 &=& \frac{\rho}{K} 
\sum_{k=1}^K \tilde{q}_{kk} {\bf y}_k^* {\bf Q}_k {\bf D}_0 {\bf Q} 
{\bf D}_\ell {\bf T} {\bf y}_k \nonumber \\
&=& 
\frac{\rho}{K} 
\sum_{k=1}^K \tilde{t}_{k} {\bf y}_k^* {\bf Q}_k {\bf D}_0 {\bf Q} 
{\bf D}_\ell {\bf T} {\bf y}_k 
+ 
\frac{\rho}{K} 
\sum_{k=1}^K ( \tilde q_{kk} - \tilde{t}_{k} )
{\bf y}_k^* {\bf Q}_k {\bf D}_0 {\bf Q} {\bf D}_\ell {\bf T} {\bf y}_k 
\nonumber \\ 
&\stackrel{\triangle}=& X'_3 + \epsilon_2 \ .
\label{eq-X3} 
\end{eqnarray} 
By Cauchy-Schwartz inequality, 
$$
\EE |\epsilon_2| \le 
\frac{\rho}{K} \sum_{k=1}^K 
\sqrt{\EE ( \tilde q_{kk} - \tilde{t}_{k} )^2 } 
\sqrt{\EE
({\bf y}_k^* {\bf Q}_k {\bf D}_0 {\bf Q} {\bf D}_\ell {\bf T} {\bf y}_k)^2} 
\ . 
$$
We have
$\EE
({\bf y}_k^* {\bf Q}_k {\bf D}_0 {\bf Q} {\bf D}_\ell {\bf T} {\bf y}_k)^2 
\le \sigma_{\max}^8  \rho^{-6} \EE \| {\bf y}_k \|^4 \le C$. 
Using in addition Lemma \ref{lm-props-Q}--(\ref{(q-t)-individuels}), 
we obtain
$$
\EE |\epsilon_2| \le \frac{C}{\sqrt{K}}  \ . 
$$
Consider $X'_3$. From \eqref{eq-qnn} and \eqref{eq-Q-lemme-inversion}, 
we have 
${\bf Q} = {\bf Q}_k - \rho \tilde{q}_{kk} {\bf Q}_k {\bf y}_k {\bf y}_k^*
{\bf Q}_k$. Hence, we can develop $X'_3$ as 
\begin{eqnarray} 
X'_3 &=&  
\frac{\rho}{K} \sum_{k=1}^K \tilde{t}_k {\bf y}_k^* {\bf Q}_k {\bf D}_0 
{\bf Q}_k {\bf D}_\ell {\bf T} {\bf y}_k 
- 
\frac{\rho^2}{K} 
\sum_{k=1}^K \tilde{t}_k \tilde q_{kk} 
{\bf y}_k^* {\bf Q}_k {\bf D}_0 {\bf Q}_k {\bf y}_k
{\bf y}_k^* {\bf Q}_k {\bf D}_\ell {\bf T} {\bf y}_k \nonumber \\
&\stackrel{\triangle}=& X_4 + X_5 \ . 
\label{eq-X'3} 
\end{eqnarray} 
Consider $X_4$. 
Notice that ${\bf y}_k$ and ${\bf Q}_k$ are independent. Therefore, by 
Lemma \ref{lm-jack}, we obtain
$$
{\bf y}_k^* {\bf Q}_k {\bf D}_0 {\bf Q}_k {\bf D}_\ell {\bf T} {\bf y}_k 
= 
\frac{1}{K} 
\tr {\bf D}_k {\bf Q}_k {\bf D}_0 {\bf Q}_k {\bf D}_\ell {\bf T} 
+ \epsilon_3 
=
\frac{1}{K} 
\tr {\bf D}_k {\bf Q} {\bf D}_0 {\bf Q} {\bf D}_\ell {\bf T} 
+ \epsilon_3 + \epsilon_4 
$$
where $\EE\epsilon_3^2 < C K^{-1}$ by Ineq.
\eqref{eq-lm-jack-simple}. Applying twice Lemma \ref{lm-props-Q}--(\ref{rank-one}) to
$\epsilon_4 = \frac{1}{K}
( \tr {\bf D}_k {\bf Q}_k {\bf D}_0 {\bf Q}_k {\bf D}_\ell {\bf T} - 
\tr {\bf D}_k {\bf Q} {\bf D}_0 {\bf Q} {\bf D}_\ell {\bf T})$ yields
$| \epsilon_4 | < C K^{-1}$.
Note in addition that 
$\sum \tilde t_k {\bf D}_k = \diag( \tr \widetilde{\bf D}_1 \widetilde{\bf T},
\ldots, \tr \widetilde{\bf D}_N \widetilde{\bf T})$. Thus, we obtain 
\begin{eqnarray} 
X_4 &=& \frac{\rho}{K^2} 
\tr \left( \sum_{k=1}^K \tilde t_k {\bf D}_k \right) 
{\bf Q} {\bf D}_0 {\bf Q} {\bf D}_\ell {\bf T} + \epsilon_5 
\nonumber \\
&=& 
X_2 + \epsilon_5 \ ,
\label{eq-X4} 
\end{eqnarray} 
where $\epsilon_5 = \epsilon_3 + \epsilon_4$, which yields
$\EE |\epsilon_5 | \le C K^{-\frac 12}$. \\ 
We now turn to $X_5$. First introduce the following random variable:
$$
\epsilon_6 = 
\tilde t_k \tilde q_{kk} 
{\bf y}_k^* {\bf Q}_k {\bf D}_0 {\bf Q}_k {\bf y}_k
{\bf y}_k^* {\bf Q}_k {\bf D}_\ell {\bf T} {\bf y}_k 
- 
\tilde t_k \tilde q_{kk} 
\left( \frac 1K \tr {\bf D}_k {\bf Q}_k {\bf D}_0 {\bf Q}_k \right) 
\left( \frac 1K \tr {\bf D}_k {\bf Q}_k {\bf D}_\ell {\bf T} \right)
$$
Then
\begin{multline*} 
|\epsilon_6 | \le 
\frac{1}{\rho^2}
{\bf y}_k^* {\bf Q}_k {\bf D}_0 {\bf Q}_k {\bf y}_k
\left| {\bf y}_k^* {\bf Q}_k {\bf D}_\ell {\bf T} {\bf y}_k 
- 
\frac 1K \tr {\bf D}_k {\bf Q}_k {\bf D}_\ell {\bf T} \right| \\
+ 
\frac{1}{\rho^2} 
\left| 
{\bf y}_k^* {\bf Q}_k {\bf D}_0 {\bf Q}_k {\bf y}_k^*
- \frac 1K \tr {\bf D}_k {\bf Q}_k {\bf D}_0 {\bf Q}_k \right|
\frac 1K \tr {\bf D}_k {\bf Q}_k {\bf D}_\ell {\bf T}  
\end{multline*} 
and one can prove that $\EE | \epsilon_6 | < C K^{-\frac 12}$ with help of 
Lemma \ref{lm-jack}, together with Cauchy-Schwarz inequality. 
In addition, we can prove with the help of Lemma \ref{lm-props-Q} that: 
\begin{eqnarray*} 
\tilde t_k \tilde q_{kk}
\left( \frac 1K \tr {\bf D}_k {\bf Q}_k {\bf D}_0 {\bf Q}_k \right) 
\left( \frac 1K \tr {\bf D}_k {\bf Q}_k {\bf D}_\ell {\bf T} \right) 
&=&  
\tilde t_k^2 
\left( \frac 1K \tr {\bf D}_k {\bf Q} {\bf D}_0 {\bf Q} \right) 
\left( \frac 1K \tr {\bf D}_k {\bf Q} {\bf D}_\ell {\bf T} \right) 
+ \epsilon_7 \\
&=& 
\tilde t_k^2 
\left( \frac 1K \tr {\bf D}_k {\bf Q} {\bf D}_0 {\bf Q} \right) 
\left( \frac 1K \tr {\bf D}_k {\bf D}_\ell {\bf T}^2 \right) 
+ \epsilon_7 + \epsilon_8 
\end{eqnarray*} 
where $\epsilon_7$ and $\epsilon_8$ are random variables satisfying
$\EE | \epsilon_7 | < C K^{-\frac 12}$ by Lemma
\ref{lm-props-Q}, and
$ \max_{k,\ell} \EE | \epsilon_8 | \le \max_{k,\ell} \sqrt{\EE
  |\epsilon_8|^2} \le C K^{-\frac 12}$ by Lemma
\ref{lm-props-generales-Q-T}--(\ref{lm-borne-trSQT}). Using the fact that
$\rho^2 \tilde{t}_k^2 = (1 + \frac 1K \tr {\bf D}_k {\bf T})^{-2}$, we
end up with
\begin{equation}
\label{eq-X5} 
X_5 = 
- \frac{\rho^2}{K} \sum_{k=1}^K \tilde t_k^2
\left( \frac 1K \tr {\bf D}_k {\bf Q} {\bf D}_0 {\bf Q} \right)
\left( \frac 1K \tr {\bf D}_k {\bf D}_\ell {\bf T}^2 \right)
+ \epsilon_9 
= 
- \sum_{k=1}^K c_{\ell k} U_k + \epsilon_9 
\end{equation} 
where $c_{\ell k}$ is given by \eqref{eq-clk}, and where 
$\EE| \epsilon_9 | < C K^{-\frac 12}$. 

Plugging Eq. (\ref{eq-X1})--(\ref{eq-X5}) into \eqref{eq-Ul}, we
end up with $U_\ell = \sum_{k=1}^K c_{\ell k} U_k + \frac 1K \tr {\bf
  D}_0 {\bf D}_\ell {\bf T}^2 + \epsilon$ with $\EE | \epsilon | < C
K^{-\frac 12}$. Step 3 is established.

\paragraph*{Step 4 : Proof of \eqref{eq-U0}} 
We rely on results of 
Section \ref{sec-proof-th-clt-bornes}, in particular on Lemma 
\ref{lm-props-A}. 

Define the following $(K+1)\times 1$ vectors:
$$
{\bf u} = [ U_k ]_{k=0}^K,\ {\bf d} = \left[ \frac 1K \tr {\bf D}_0 {\bf D}_k {\bf T}^2 
\right]_{k = 0}^K,\  {\bs \epsilon} = [ \epsilon_k ]_{k=0}^K\ ,
$$
where the $U_k$'s and $\epsilon_k$'s are defined in 
\eqref{eq-syst-Ul}. Recall the definition of the $c_{\ell k}$'s for $0\le \ell \le K$ and 
$1\le k\le K$, define $c_{\ell\,0}=0$ for $0\le \ell\le K$ and consider the $(K+1)\times (K+1)$ matrix
${\bf C} = [ c_{\ell k} ]_{\ell, k = 0}^K$.

With these notations, System \eqref{eq-syst-Ul} writes 
\begin{equation}
\label{eq-u=Cu}
\left( {\bf I}_{K+1} - {\bf C} \right) {\bf u} = {\bf d} + {\bs \epsilon} \ . 
\end{equation} 
Let ${\bs \alpha} = \frac 1K \tr\, {\bf D}_0^2 {\bf T}^2$ and 
${\bs \beta} = ( 1 + \frac 1K \tr\,{\bf D}_0 {\bf T} )^2$. We have in particular 
$$
{\bf d} = \left[\begin{array}{c} {\bs \alpha} \\ {\bf g} \end{array}\right],
\quad
{\bf C} = \left[\begin{array}{cc}
0 & 
\frac 1K {\bf g}^T {\bs \Delta}^{-1} \\
{\bf 0} & {\bf A}^T 
\end{array}\right] 
$$
(recall that ${\bf A}$, ${\bs \Delta}$ and ${\bf g}$ are defined in
the statement of Theorem \ref{th-clt}). \\  
Consider a square matrix ${\bf X}$ which first column is equal to 
$[ 1, 0, \ldots, 0 ]^\T$, and partition ${\bf X}$ as 
${\bf X} = \left[\begin{array}{cc} 1 & {\bf x}_{01}^T \\
{\bf 0} & {\bf X}_{11} \end{array}\right]$.
Recall that the inverse of ${\bf X}$ exists if and only if 
${\bf X}_{11}^{-1}$ exists, and in this case 
the first row $[ {\bf X}^{-1} ]_0$ of ${\bf X}^{-1}$ is given by
$$
\left[ {\bf X}^{-1}\right]_0 = 
\left[ 1 \ \ -{\bf x}_{01}^T {\bf X}_{11}^{-1} \right]  
$$
(see for instance \cite{hor-joh-livre94}). 
We now apply these results to the system \eqref{eq-u=Cu}.
Due to \eqref{eq-u=Cu}, $U_0$ can be expressed as 
$$
U_0 = [ ( {\bf I} -
{\bf C} )^{-1} ]_0 ( {\bf d} + {\bs \epsilon})\ .
$$ 
By Lemma \ref{lm-props-A}--(\ref{lm-props-A-inversible}), 
$({\bf I}_K - {\bf A}^T)^{-1}$ exists hence $({\bf I} - {\bf C})^{-1}$ exists, 
$$
\left[ \left({\bf I}_{K+1} - {\bf C}\right)^{-1}\right]_0 = 
\left[ 1 \ \ \frac 1K {\bf g}^T {\bs \Delta}^{-1}
({\bf I}_K - {\bf A}^T)^{-1} \right] \ , 
$$
and
$$
U_0 = 
 {\bs \alpha} + 
\frac 1K {\bf g}^T {\bs \Delta}^{-1}
\left({\bf I} - {\bf A}^T\right)^{-1} {\bf g} 
+ 
\epsilon_0 + \frac 1K {\bf g}^T {\bs \Delta}^{-1}
\left({\bf I} - {\bf A}^T\right)^{-1} {\bs \epsilon}' 
$$
with ${\bs \epsilon}' = [ \epsilon_1, \ldots, \epsilon_K ]^T$. Gathering the estimates 
of Section \ref{sec-proof-th-clt-bornes} together with the fact that $\|  \EE {\bs \epsilon}  \|_\infty 
\le C K^{-\frac 12}$, we get \eqref{eq-U0}. Step 4 is established, so is Theorem \ref{th-clt}.

\appendix
\subsection{Proof of Lemma \ref{lm-props-generales-Q-T}}
\label{anx-proof-lm-props-generales-Q-T} 
Let us establish \eqref{eq-bornes-T}. The lower bound immediately follows from 
the representation
$$
t_n = \frac{1}
{\rho + \frac 1K \sum_{k=1}^K 
\frac{\sigma^2_{nk}}
{ 1 + \frac 1K \sum_{\ell=1}^N \sigma^2_{\ell k} t_{\ell} } } 
 \stackrel{(a)}{\geq} \frac{1}{\rho + \sigma_{\max}^2} 
$$  
where $(a)$ follows from {\bf A2} and $t_{\ell}(-\rho)\ge 0$. The
upper bound requires an extra argument: As proved in \cite[Theorem
2.4]{hac-lou-naj-aap07}, the $t_{n}$'s are Stieltjes transforms of
probability measures supported by $\RR_+$, i.e. there exists a
probability measure $\mu_{n}$ over $\RR_+$ such that $t_{n}(z) =
\int \frac{\mu_{n}(dt)}{t-z} $. Thus 
$$
t_n(-\rho) = \int_0^\infty \frac{\mu_n(dt)}{t+\rho}  \leq \frac 1 \rho \ ,
$$
and \eqref{eq-bornes-T} is proved.

We now briefly justify \eqref{eq-borne-trSQT}.  
We have $\EE \left| \tr\, {\bf S} ( {\bf Q} - {\bf T} ) \right|^2  
= 
\EE \left| \tr {\bf S} ( {\bf Q} - \EE{\bf Q} ) \right|^2  
+ \left| \tr {\bf S} ( \EE{\bf Q} - {\bf T} ) \right|^2$. 
In \cite[Lemma 6.3]{hac-lou-naj-(clt)-(sub)aap07} it is stated that 
$\sup_K \EE \left| \tr {\bf S} ( {\bf Q} - \EE{\bf Q} ) \right|^2 < \infty$. 
Furthermore, 
in the proof of \cite[Theorem 3.3]{hac-lou-naj-(clt)-(sub)aap07} it is 
shown that $\sup_K K \| \EE{\bf Q} - {\bf T} \| < \infty$, hence
$\left| \tr {\bf S} ( \EE{\bf Q} - {\bf T} ) \right| \le  
K \| {\bf S} ( \EE{\bf Q} - {\bf T} ) \| \le K \| \EE{\bf Q} - {\bf T} \| 
\| {\bf S} \| < \infty$ by Lemma 
\ref{lm-matrix-analysis}--(\ref{ineq-normes}). The result follows. 

\subsection{Proof of Corollary \ref{cor-clt-separable}} 
\label{anx-proof-cor-clt-separable} 
Recall that in the separable case, ${\bf D}_k = \tilde{d}_k {\bf D}$ 
and $\widetilde{\bf D}_n = d_n \widetilde{\bf D}$. 
Let $\tilde{\bf d}$ be the $K\times 1$ vector 
$\tilde{\bf d} = [ \tilde d_k ]_{k=1}^K$.  
In the separable case, Eq. \eqref{eq-theta} is written
\begin{equation}
\label{eq-theta-separable} 
\frac{\Theta^2}{\tilde d_0^2} = 
\frac{1}{K \tilde d_0^2} 
{\bf g}^T ( {\bf I} - {\bf A})^{-1} {\bs \Delta}^{-1} {\bf g}
+ 
\gamma ( \EE| W_{10} |^4 - 1 )\ ,
\end{equation} 
where $\gamma$ is defined in statement of the corollary. 
Here, vector ${\bf g}$ and matrix ${\bf A}$ are given by 
$$
{\bf g} = \gamma \tilde{d}_0 \tilde{\bf d} \quad \text{and} \quad 
{\bf A} =
\left[
\frac 1K \frac{\frac{1}{K} \tr {\bf D}_{\ell} {\bf D}_{m} {\bf T}^2}
{\left(1 + \frac{1}{K} \tr {\bf D}_{\ell} {\bf T} \right)^2}
\right]_{\ell,m=1}^K = 
\frac{\gamma}{K} {\bs \Delta}^{-1} \tilde{\bf d} \tilde{\bf d}^T \ . 
$$
By the matrix inversion lemma \cite{hor-joh-livre94}, we have 
\begin{eqnarray*}
\frac{1}{K \tilde d_0^2}
{\bf g}^T ( {\bf I} - {\bf A})^{-1} {\bs \Delta}^{-1} {\bf g}
&=& 
\frac{\gamma^2}{K} 
\tilde{\bf d}^T 
\left( {\bs \Delta} - \frac{\gamma}{K} \tilde{\bf d} \tilde{\bf d}^T
\right)^{-1} \tilde{\bf d} \\
&=&
\frac{\gamma^2}{K} \tilde{\bf d}^T 
\left( {\bs \Delta}^{-1} + \frac{\gamma}{K} 
\frac{1}{1 - 
\frac{\gamma}{K} \tilde{\bf d}^T {\bs \Delta}^{-1} \tilde{\bf d}} 
{\bs \Delta}^{-1} \tilde{\bf d} \tilde{\bf d}^T {\bs \Delta}^{-1} 
\right) 
\tilde{\bf d}\ .
\end{eqnarray*} 
Noticing that 
$$
\frac{1}{K} \tilde{\bf d}^T {\bs \Delta}^{-1} \tilde{\bf d} 
= 
\frac 1K \sum_{k=1}^K 
\frac{\tilde d_k^2}{\left( 1 + \frac 1K \tr {\bf D}_k {\bf T} \right)^2} 
= 
\frac{\rho^2}{K} \sum_{k=1}^K \tilde d_k^2 \tilde t_k^2 = 
\rho^2 \tilde\gamma \ ,
$$
we obtain
$$
\frac{1}{K \tilde d_0^2}
{\bf g}^T ( {\bf I} - {\bf A})^{-1} {\bs \Delta}^{-1} {\bf g}
= 
\gamma \frac{\rho^2 \gamma \tilde\gamma}{1 - \rho^2 \gamma\tilde\gamma} \ .
$$
Plugging this equation into \eqref{eq-theta-separable}, we obtain 
\eqref{eq-omega-separable}. 


\subsection{Proof of Lemma \ref{lm-props-Q}}
The proof of Part \ref{rank-one} can be found in 
\cite[Proof of Lemma 6.3]{hac-lou-naj-(clt)-(sub)aap07} (see also 
\cite[Lemma 2.6]{sil-bai-jma95}). Let us prove Part \ref{(q-t)-individuels}.
We have from Equations \eqref{eq-systeme-T} and \eqref{eq-qnn}
\begin{eqnarray*}
| q_{nn}(-\rho) - t_{n}(-\rho) | &=& 
\frac{1}
{\rho(1+ \frac 1K \tr \widetilde{\bf D}_n \widetilde{\bf T}) 
(1+{\bs \xi}_n \widetilde{\bf Q}_n {\bs \xi}_n^*)} 
\left| 
{\bs \xi}_n \widetilde{\bf Q}_n {\bs \xi}_n^* - 
\frac 1K \tr \widetilde{\bf D}_n \widetilde{\bf T} 
\right|  \\
&\le& 
\frac 1 \rho 
\left| 
{\bs \xi}_n \widetilde{\bf Q}_n {\bs \xi}_n^* - 
\frac 1K \tr \widetilde{\bf D}_n \widetilde{\bf T}  \right| \ .
\end{eqnarray*}
Hence,
$$
\EE ( q_{nn} - t_{n} )^2 
\quad \le \quad
\frac{2}{\rho} 
\EE \left( {\bs \xi}_n \widetilde{\bf Q}_n {\bs \xi}_n^* - 
\frac 1K \tr \widetilde{\bf D}_n \widetilde{\bf Q}  \right)^2 
+ 
\frac{2}{\rho K^2} 
\EE \left( 
\tr \widetilde{\bf D}_n ( \widetilde{\bf Q} - \widetilde{\bf T} )
\right)^2 
\quad\le\quad \frac{C}{K} 
$$
by Lemma \ref{lm-jack} and Lemma 
\ref{lm-props-generales-Q-T}--(\ref{lm-borne-trSQT}), which proves 
\eqref{eq-E(q-t)}. 
\bibliographystyle{IEEEbib}
\bibliography{BSTLabbrev,bibli}

\end{document}